\documentclass[useAMS,usenatbib,fleqn]{mnras}
\usepackage{multirow}
\usepackage{graphicx}
\usepackage{color,soul}
\usepackage{epsfig}
\usepackage{amssymb,amsmath}
\usepackage{aas_macros}
\usepackage{fixltx2e}
\DeclareMathOperator{\sinc}{sinc}

\usepackage[T1]{fontenc}
\defcitealias{EHT5}{EHT5}

\title[BH Internal Faraday Rotation]{
Decomposing the Internal Faraday Rotation of Black Hole Accretion Flows
}

\author[Ricarte et al.]{Angelo Ricarte$^{1,2}$, Ben S.~Prather$^{3}$, George N.~Wong$^{3}$, Ramesh Narayan$^{1,2}$, 
\newauthor
Charles Gammie$^{3,4}$, and Michael D.~Johnson$^{1,2}$
\\
$^{1}$ Center for Astrophysics | Harvard \& Smithsonian, 60 Garden Street, Cambridge, MA 02138, USA \\
$^{2}$ Black Hole Initiative, 20 Garden Street, Cambridge, MA 02138, USA \\
$^{3}$ Department of Physics, University of Illinois at Urbana–Champaign, 1110 West Green Street, Urbana, IL 61801, USA \\
$^{4}$ Department of Astronomy, University of Illinois at Urbana–Champaign, 1002 West Green Street, Urbana, IL 61801, USA
}

\date{\today}

\begin{document}
\pagerange{\pageref{firstpage}--\pageref{lastpage}} \pubyear{2020}
\maketitle

\begin{abstract}
Faraday rotation has been seen at millimeter wavelengths in several low luminosity active galactic nuclei, including Event Horizon Telescope (EHT) targets M87* and Sgr A*.  The observed rotation measure (RM) probes the density, magnetic field, and temperature of material integrated along the line of sight.  To better understand how accretion disc conditions are reflected in the RM, we perform polarized radiative transfer calculations using a set of general relativistic magneto-hydrodynamic (GRMHD) simulations appropriate for M87*.  We find that in spatially resolved millimetre wavelength images on event horizon scales, the RM can vary by orders of magnitude and even flip sign.  The observational consequences of this spatial structure include significant time-variability, sign-flips, and non-$\lambda^2$ evolution of the polarization plane.  For some models, we find that internal rotation measure can cause significant bandwidth depolarization even across the relatively narrow fractional bandwidths observed by the EHT.  We decompose the linearly polarized emission in these models based on their RM and find that emission in front of the mid-plane can exhibit orders of magnitude less Faraday rotation than emission originating from behind the mid-plane or within the photon ring.  We confirm that the spatially unresolved (i.e., image integrated) RM is a poor predictor of the accretion rate, with substantial scatter stemming from time variability and inclination effects.  Models can be constrained with repeated observations to characterise time variability and the degree of non-$\lambda^2$ evolution of the polarization plane.
\end{abstract}


\section{Introduction}
\label{sec:introduction}

\vspace{20pt}

Using a network of millimetre telescopes around the world, the Event Horizon Telescope (EHT) has recently produced the first images of a black hole (BH) accretion flow \citep{EHT1,EHT2,EHT3,EHT4,EHT5,EHT6}.  These images resolve the ``shadow'' of the supermassive BH M87*, corresponding to rays that begin on its event horizon, providing new constraints on the properties of the BH and its accretion disc.  

In \citet{EHT5}, henceforth \citetalias{EHT5}, a library of general relativistic magneto-hydrodynamic (GRMHD) simulations was produced to compare to the EHT images, exploring three fundamental quantities.  The first is the strength of the magnetic field: models that accumulate strong magnetic flux around the BH are able to counteract the ram pressure of in-falling gas with magnetic pressure, resulting in what is termed a Magnetically Arrested Disc (MAD) \citep{Igumenshchev+2003,Narayan+2003,Chael+2019}.  In contrast, the weaker and more turbulent magnetic fields in Standard And Normal Evolution (SANE) models have a weaker effect on the gas dynamics of the disc \citep{Narayan+2012,Sadowski+2013,Ryan+2018}.  The second quantity is the BH's angular momentum, described by the dimensionless spin parameter $a \equiv Jc/(GM_\bullet^2) \in (-1,1)$, where negative values correspond to counter-rotating accretion discs.  Mass and spin are the only properties intrinsic to an astrophysical BH, but BH spins are constrained much more loosely than their masses.  Most of our understanding of supermassive BH spin evolution originates from theory \citep[e.g.,][]{Bardeen&Wagoner1969,Thorne1974,Gammie+2004,King+2008,Volonteri+2013}.  The third quantity explored in this work is $R_\mathrm{high}$, one parameterisation of the relative temperatures of electrons and ions in the plasma.  Such a prescription is necessary because the mean free path near the event horizon is so large, the two populations depart from thermal equilibrium and divide into a two-temperature plasma \citep{Shapiro+1976,Rees+1982,Narayan&Yi1995b,Sadowski+2017,Ryan+2018}.  By combining EHT imaging with other multi-wavelength constraints such as the X-ray flux and jet power, \citetalias{EHT5} rule out all but 19/60 possible models which span these three variables.  In particular, all $a = 0$ models are excluded.

EHT constraints on M87* thus far have only considered total intensity (Stokes $I$), when in fact polarized visibilities have been obtained (Stokes $I$, $Q$, $U$, and $V$), but have not been published.  Hence, current constraints have only utilized 1/4 of the measured information.  Synchrotron emission, the emission mechanism at millimetre wavelengths, has a near-unity intrinsic polarization fraction \citep{LeRoux1961,Bromley+2001,Broderick&Blandford2003}.  Polarimetric imaging is consequently predicted to tightly constrain accretion models, which differ substantially in their linear polarization fractions as well as their morphologies \citep{Palumbo+2020}.  Although an image has not yet been constructed for Sgr A*, strong and variable linear polarization has been observed for this source. Previous very long baseline interferometry (VLBI) measurements have revealed a partially ordered magnetic field structure at its centre \citep{Johnson+2015}. 

As polarized radiation travels through a magnetized plasma, it is transformed by effects sensitive to the local density, temperature, and magnetic field.  Faraday rotation turns the electric vector position angle (EVPA) of linearly polarized emission, and Faraday conversion exchanges linearly and circularly polarized radiation \citep{Sazonov1969,Rybicki&Lightman1986,Melrose1997,Shcherbakov2008}.  Future analyses with the EHT will be able to further distinguish models in the time and frequency domains \citep[e.g.,][]{Broderick&Loeb2006a,Broderick&Loeb2006b,Roelofs+2017,Medeiros+2018}.  It is therefore timely for us to better understand time and frequency dependent effects which may help us constrain accretion models.

One such polarimetric observable is the rotation measure (RM),  defined by the change in the EVPA, $\chi$, as a function of the change of wavelength squared, 

\begin{equation}
    \mathrm{RM} = \frac{\chi_2 - \chi_1}{\lambda^2_2 - \lambda^2_1}, \label{eqn:rm_definition}
\end{equation}

\noindent where the subscripts 1 and 2 denote two different wavelengths.  RM probes Faraday rotation, and has been used in a variety of contexts to infer magnetic field properties \citep[e.g.,][]{Zavala&Taylor2004,Brentjens&deBruyn2005,Frick+2011,Pasetto+2018,Agudo+2018}.  For a source of polarized emission that is entirely behind a Faraday rotating medium, the RM can be written as an integral of plasma properties along the line-of-sight via

\begin{equation}
    \mathrm{RM} = 8.1\times10^5 \ \mathrm{rad}\; \mathrm{m}^{-2} \int_\mathrm{source}^\mathrm{observer} f_\mathrm{rel}(\Theta_e)\frac{n_e}{1 \ \mathrm{cm}^{-3}} \frac{B_{||}}{G}\frac{ds}{\mathrm{pc}},
    \label{eqn:rm_physics}
\end{equation}

\noindent where $n_e$ is the electron number density, $B_{||}$ is the parallel component of the magnetic field, and $f_\mathrm{rel}$ is a correction term suppressing Faraday rotation at relativistic temperatures \citep{Gardner&Whiteoak1966}.  At relativistically hot temperatures, $f_\mathrm{rel}(\Theta_e) \approx \log(\Theta_e)/(2\Theta_e^2)$, whereas at sub-relativistic temperatures, $f_\mathrm{rel}$ asymptotes to 1.  Here $\Theta_e \equiv k_BT_e/m_ec^2$, $k_B$ is the Boltzmann constant, $T_e$ is the electron temperature, $m_e$ is the electron rest mass, and $c$ is the speed of light \citep{Jones&Odell1977}.

As seen from equation \ref{eqn:rm_physics}, the RM directly traces the electron temperature, number density, and magnetic field along the line of sight.  Among the models consistent with the EHT data of M87*, all of these quantities can vary by orders of magnitude.  Note that while RM scales as $\mathrm{RM} \propto n_eB$, the power emitted by synchrotron emission scales as $P \propto n_eB^2$ \citep{Rybicki&Lightman1986}.  In principle, this could allow RM to break a degeneracy that exists between $n_e$ and $B$ based on total intensity alone.

RMs have been measured for the two main EHT targets Sgr A* and M87*, as well as a handful of other low-luminosity AGN.  For Sgr A*, $\mathrm{RM} = -5\times10^5 \ \mathrm{rad} \; \mathrm{m}^{-2}$ \citep{Bower+2003,Marrone+2007,Bower+2018} while $-7 \times 10^4 \ \mathrm{rad} \; \mathrm{m}^{-2}$ has been measured a few arcseconds away \citep{Eatough+2013}.  For M87*, an upper limit of $|\mathrm{RM}| < 7\times10^5 \ \mathrm{rad}\;\mathrm{m}^{-2}$ was measured using the Submillimeter Array \citep{Kuo+2014}.  3C 84 has an RM of $10^{5-6} \ \mathrm{rad}\;\mathrm{m}^{-2}$ \citep{Kim+2019} at 43 GHz and $9 \times 10^{5} \ \mathrm{rad}\;\mathrm{m}^{-2}$ at 230 GHz \citep{Plambeck+2014}.  Similarly, 3C 273 has an RM of $5 \times 10^{5} \ \mathrm{rad}\;\mathrm{m}^{-2}$ at 230 GHz \citep{Hovatta+2019}.  Neither linear polarization nor RM could be measured for the low-luminosity AGNs M81 and M84, which might imply significant  scrambling \citep{Brunthaler+2001,Bower+2017}.

These measurements have been used to constrain the accretion rates of EHT targets Sgr A* and M87* by assuming simple analytic models describing the accretion flow that can be input into equation \ref{eqn:rm_physics} \citep{Marrone+2006}.  The accretion rates of Sgr A* and M87* have thus been constrained to $<10^{-6} \ M_\odot \; \mathrm{yr}$ (although it may be much lower) \citep{Agol2000,Quataert&Gruzinov2000b,Marrone+2006} and $<9.2\times10^{-4} \ M_\odot \; \mathrm{yr}$ \citep{Kuo+2014,Li+2016} respectively.  However, there are important effects that complicate the simple scenario implicitly assumed by equation \ref{eqn:rm_physics}, where a Faraday rotator sits entirely between a source and our line-of-sight \citep[see also]{Broderick&McKinney2010}.  Most importantly, for BH accretion flows, Faraday rotation and emission occur co-spatially, such that along a given geodesic, not all photons are Faraday rotated by the same material.  In addition, general relativity (GR) complicates a photon's trajectory and can modify its polarization properties by parallel transport alone.  This is especially true of emission near the photon ring, where null geodesics can pass through the accretion flow multiple times in different directions, leading to interesting polarization signatures \citep{Johnson+2019,Himwich+2020}.  Finally, neither the emission nor the Faraday rotation can be assumed to behave in a spatially uniform way, and the magnetic field may switch sign, especially in turbulent SANE discs. 

\citet{Moscibrodzka+2017} produced the first polarized model images of M87* at millimetre wavelengths from ray traced GRMHD simulations.  Using a set of SANE models, they determined that Faraday rotation can be strong enough to spatially scramble the polarization from the counter-jet, which must pass through a larger Faraday depth than the forward-jet on the way to the observer.  Their analysis also revealed a significant inclination dependence of the RM, such that even very high accretion rate models could satisfy RM constraints when viewed face-on. \citet{Jimenez-Rosales&Dexter2018} determined that high-accretion rate models are disfavoured, since this scrambling too strongly depolarizes the emission.  \citet{Tsunetoe+2020} began to explore the spin dependence and favoured $a=0.9$ models, although an expanded parameter survey is warranted.

In this work, we perform a more comprehensive survey of RM for 7  models consistent with \citetalias{EHT5}, chosen to bracket the allowed parameter space.  We consider variations as a function of time and inclination, and develop new techniques to model Faraday effects in some models.  In \S\ref{sec:methodology}, we describe the GRMHD simulations we use as a starting point, the radiative transfer calculations in post-processing, and a novel Taylor expansion model for treating internal Faraday rotation.  In \S\ref{sec:results}, we describe our results.  This includes an exploration of strong spatial variations in RM, RM distribution functions, RM as a measure of accretion rate, inclination dependence, the degree of non-$\lambda^2$ behaviour among models, and a case study of time variability.  We discuss our results in \S\ref{sec:discussion}, and end with a summary and conclusion in \S\ref{sec:conclusion}.

\section{Methodology}
\label{sec:methodology}

We begin with a set of GRMHD simulations that are consistent with \citetalias{EHT5}.  We then use {\tt ipole}\footnote{\url{https://github.com/moscibrodzka/ipole/}} \citep{Moscibrodzka&Gammie2018} to perform polarized radiative transfer calculations, specifying electron properties in this step.  Finally, using a first order Taylor expansion, we create a model of the polarized image in order to capture frequency dependent effects and compute rotation measures.

\subsection{GRMHD Simulations and Radiative Transfer}
\label{sec:ipole}

\begin{figure*}
    \centering
    \includegraphics[width=\textwidth]{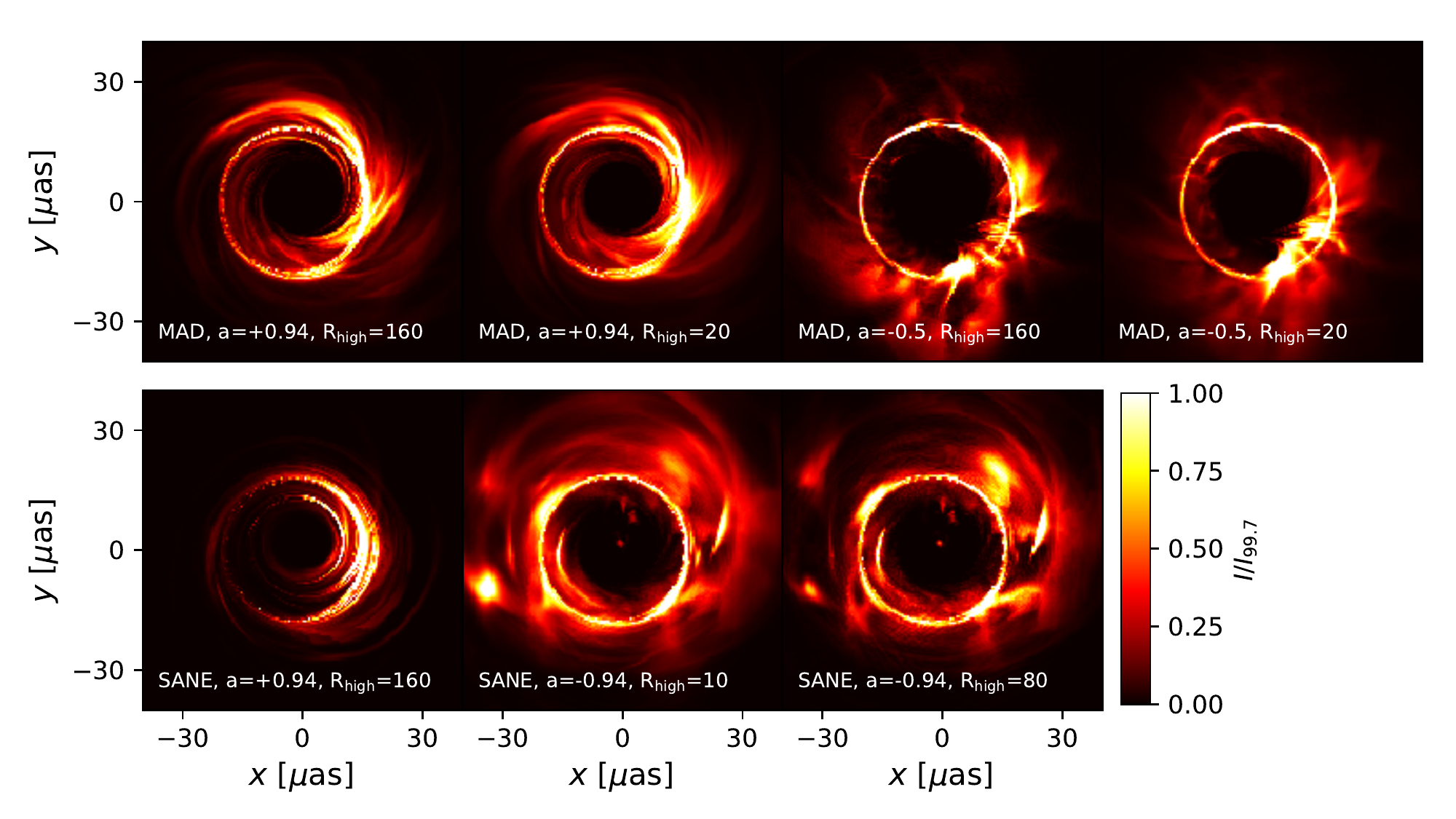}
    \caption{Total intensity images of the 7 models considered in this work, taken at the final snapshot of each GRMHD simulation, which occurs at time $t=10000 \ GM_\bullet/c^3$.  Models are tilted by 17$^\circ$ from face-on towards the top of these images and in subsequent figures.  To help visualise low surface brightness features, intensity is scaled with respect to the intensity of the pixel in the 99.7th percentile, $I_{99.7}$, saturating 0.3 per cent of the pixels.  \label{fig:intensity_examples}}
\end{figure*}

In this work, we study 7 models for M87* in the EHT GRMHD simulation library, each of which is consistent with all of the observational constraints considered in \citetalias{EHT5}.  The properties of these simulations, all performed with {\tt iharm} \citep{Gammie+2003}, are listed in Table \ref{tab:models}.  Both SANE and MAD models are considered, while values of $a$ and $R_\mathrm{high}$ are chosen to bracket the allowed values in \citetalias{EHT5}.  These models are described in more detail in \citetalias{EHT5}, and the SANE $a=+0.94$ model is included in the GRMHD code comparison project of \citet{Porth+2019}.  

As defined by \citet{Moscibrodzka+2016}, $R_\mathrm{high}$ prescribes the electron temperature via

\begin{equation}
    \frac{T_i}{T_e} = R_\mathrm{high}\frac{\beta_p^2}{1+\beta_p^2} + \frac{1}{1+\beta_p^2}
\end{equation}

\noindent where $T_i$ and $T_e$ are the ion and electron temperatures respectively, and $\beta_p$ is the ratio of gas to magnetic pressure.  $T_i$ is determined by the GRMHD simulation.  In the mid-plane, where $\beta_p$ is high, $T_e \rightarrow T_i/R_\mathrm{high}$, since turbulent plasma models reveal that heating preferentially affects ions, which then cannot efficiently transfer energy to electrons \citep{Rees+1982,Narayan&Yi1995,Quataert&Gruzinov1999,Howes2010,Kawazura+2019}.  Consequently, increasing $R_\mathrm{high}$ has the effect of decreasing the emission from and increasing the Faraday rotation within the mid-plane.  As a result of decreasing mid-plane emission, the accretion rate must also be scaled upwards to obtain the correct total intensity for M87*.  As we shall show, both of these effects have important implications for the RM.  

Note that among these 7 models, there are only 4 unique GRMHD simulations, since $R_\mathrm{high}$ only affects the radiative transfer in post-processing.  In each of these simulations, the angular momentum of the disc is either perfectly aligned (denoted by positive spin) or anti-aligned (denoted by negative spin) with that of the BH, although misaligned discs remain an active area of research \citep[e.g.,][]{Fragile+2007,Liska+2020,Chatterjee+2020}.  MAD simulations are run with a $384\times 192\times 192$ grid with a maximum radius of $10^3 \ GM_\bullet/c^2$, while SANE simulations are run with a $288\times 128\times 128$ grid and a maximum radius of $50 \ GM_\bullet/c^2$.  However, we find that these models only exhibit inflow equilibrium within a radius of approximately $20 \; GM_\bullet/c^2$.  Throughout this work, we restrict the integration of our radiative transfer equations to within this radius.  Fortunately, this limitation actually has negligible effect on our results for face-on inclinations ($i \lesssim 30^\circ$), as we explore in Appendix \ref{sec:50M}.  This is because the funnel region is largely evacuated in these simulations and does not contribute to Faraday rotation.  However, restricting our calculations to $\leq 20 \; GM_\bullet/c^2$ may lead us to underestimate the total Faraday rotation at inclinations $\gtrsim 30^\circ$, due to material that may exist in more distant, unequilibrated regions of the simulations.

\begin{table}
\centering
\begin{tabular}{lll}
\hline
Magnetic Field State & $a$ & $\mathrm{R}_\mathrm{high}$ \\
\hline
MAD            & +0.94      & 160               \\
MAD            & +0.94      & 20                \\
MAD            & -0.5       & 160               \\
MAD            & -0.5       & 20                \\
SANE           & +0.94      & 160               \\
SANE           & -0.94      & 80                \\
SANE           & -0.94      & 10                \\
\hline
\end{tabular}
\caption{Parameters of the 7 models considered in this paper.  Each of these models passes all metrics considered in \citetalias{EHT5} and are chosen to bracket the allowed parameter space.
\label{tab:models}}
\end{table}

We create polarized ray-traced images using {\tt ipole} \citep{Moscibrodzka&Gammie2018}, which first solves the null geodesic equation backwards from the image plane, then integrates forward the radiative transfer equations for the 4 Stokes parameters, $\{I, Q, U, V\}$.  Here, $I$ is the total intensity, $\sqrt{Q^2+U^2}$ and $\frac{1}{2}\arg(Q+iU)$ are the linearly polarized intensity and EVPA respectively, and $V$ is the circularly polarized intensity.  The radiative transfer equations account for synchrotron emission, synchrotron self-absorption, Faraday rotation, and Faraday conversion.  Radiative transfer coefficients follow \citet{Dexter2016} for a thermal electron distribution function, with a slight modification to $\rho_{\nu,V}$, the coefficient responsible for Faraday rotation.  As also discussed in \citet{Dexter+2020}, minor modifications are needed to ensure continuous and accurate behaviour at low temperature and frequency.  Following \citet{Shcherbakov2008}, we set

\begin{equation}
    \rho_{\nu,V} = \frac{2ne^2\nu_B}{m_e c\nu^2}\frac{K_0(\Theta_e^{-1})}{K_2(\Theta_e^{-1})} \cos(\Theta_e) g(X),
\end{equation}

\noindent where $K_0$ and $K_2$ are modified Bessel functions of the second kind, $e$ is the electron charge, $m_e$ is the electron mass, $\nu_B=eB/2\pi m_e c$, $X=\left[\frac{3}{2\sqrt{2}} 10^{-3} \frac{\nu}{\nu_c}\right]^{-1/2}$, and $g(X) = 1 - 0.11 \ln(1+0.035X)$ for cyclotron frequency $\nu_c$.

For each model, we create images for 11 snapshots spanning the last quarter of the corresponding GRMHD run, corresponding to times $t/(GM_\bullet/c^3) \in [7500, 10000]$, a duration of 880 days for M87*.  For each snapshot, we study 5 inclinations, $i \in \{5^\circ,17^\circ,30^\circ,60^\circ,90^\circ\}$.\footnote{Throughout this paper, for models with positive spin, we actually compute images for $i = 180^\circ - i_\mathrm{written}$, as in \citetalias{EHT5}.}  Then, for each snapshot and inclination, we create 6 polarized images.  We sample 3 frequencies, 226.999, 227.000 and 227.001 GHz, to construct a frequency-dependent model of the image, detailed in the following section.  We find that we must adopt extremely small differences in frequency in order to resolve the RM on geodesics that have high Faraday depth.  Otherwise, the RM in some geodesics can be underestimated due to the ``n$\pi$ degeneracy:'' the EVPA may rotate so rapidly within that multiple rotations are missed.  Then, for each of these 3 frequencies, we create 2 separate images that only include emission from either the positive or negative $z$ domains, where the $z$-axis is oriented parallel to the black hole spin (and perpendicular to the disc in these models).  

Separate images including only the top and bottom halves of the emission are helpful for modelling Faraday effects that are often significantly different between the front and back sides of the emitting region.  In these models, emission behind the mid-plane often experiences orders of magnitude more Faraday rotation than emission anterior to it, since it must pass through the relatively cold mid-plane \citep{Moscibrodzka+2017}.  This split remains helpful at high inclinations, since some emission is lensed to the opposite side of the image.  A complete image is made by summing together the individual images made with only the near- and far-side emission.  Note that when emission from only one side is included, absorption and Faraday effects from both sides remain included.

For all images, we adopt a black hole mass of $6.2\times 10^9 \ M_\odot$  and a distance of 16.9 Mpc \citep{Gebhardt+2011}, for consistency with \citetalias{EHT5}.  Gas densities are scaled such that the average image produces an intensity of 0.5 Jy at an inclination of 17$^\circ$, which is the most likely inclination at which we are viewing M87* based on its larger scale jet \citep{Walker+2018}.  The same gas density scalings are used for models at different inclinations, although their average fluxes can depart from 0.5 Jy.  We find that the total intensities of images created at an inclination of $90^\circ$ tend to be approximately a factor of 2 larger than those created with an inclination of $17^\circ$.  Each image is created at $0.5 \ \mu$as pixel resolution, with a 160 $\mu$as field of view, a factor of 2 finer angular resolution than employed in \citetalias{EHT5}.

To summarise, we consider 7 models that are consistent with EHT observations of M87*.  For each model, we create images for 11 snapshots, 5 inclinations, 3 frequencies, and the 2 sides of the accretion flow.  We also create additional images to better resolve frequency-dependent (\S\ref{sec:taylor_expansion}) and time-dependent  (\S\ref{sec:case_study}) phenomena.  In Figure \ref{fig:intensity_examples}, we plot total intensity images of the 7 models during their final snapshot at $17^\circ$ inclination.  To better visualise low surface brightness features, we intentionally saturate 0.3 per cent of the pixels in this and all subsequent visualisations.  In all of the images presented in this work, the forward-jet points straight up, and the material on the right side of the image is moving towards the observer. 

\subsection{Modelling the Polarized Image}
\label{sec:taylor_expansion}

\begin{figure*}
  \centering
  \includegraphics[width=\textwidth]{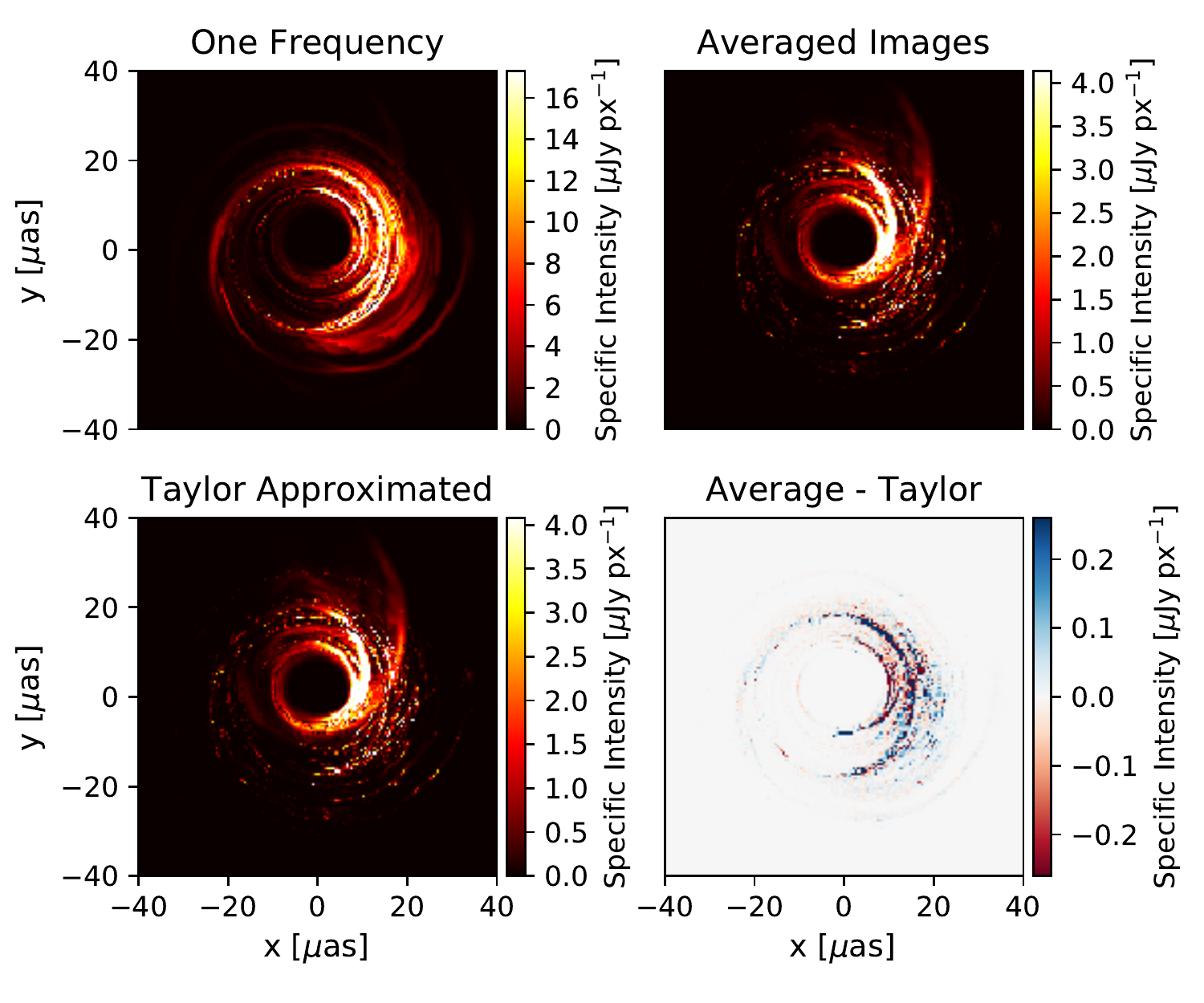}
  \caption{Images of total linearly polarized intensity ($\sqrt{Q^2+U^2}$) for a case where bandwidth depolarization is important, the SANE, $a=+0.94$, $\mathrm{R}_\mathrm{high}=160$ model, centred at 228 GHz.  In the top left, we plot an image at a single frequency, while in the top right, we plot the average of 255 images across a 2 GHz bandwidth.  As further explored in Figure \ref{fig:taylor_expansion_example_breakdown}, bandwidth depolarization suppresses the contribution of the counter-jet.  In the bottom left, we plot the result of our Taylor expansion model based on  6 images.  Its residual with respect to the 255 averaged images is shown in the bottom right.  The Taylor expansion model successfully depolarizes the counter-jet without depolarizing the forward-jet.  Since we model the image with two zones, the largest discrepancies occur in and within the photon ring, where geodesics cross the mid-plane multiple times. \label{fig:taylor_expansion_example}}
\end{figure*}

Here, we introduce a modelling formalism for the polarized image as a function of frequency.  For a given image snapshot, we create a two-zone (front and back half) model of the radiation in each pixel.  Our model is constructed as follows:

\begin{itemize}
    \item Treating the front and back halves of the emitting region separately, we first convert the Stokes parameters $\{I, Q, U, V\}$ into their counterparts on the Poincare sphere, $\{I, N, \phi, \psi\}$ \citep[following the notation of][]{Shcherbakov+2012}.  These variables have well-behaved Taylor series expansions, and are further discussed in Appendix \ref{sec:spherical_stokes}.
    \item Using images at three frequencies, we then compute the first and second derivatives of the spherical Stokes parameters.  Each derivative is computed in the variable that best approximates expected physical dependencies:  $I$ and $N$ are differentiated in $\nu$, $\phi$ is differentiated in $\lambda^2$, and $\psi$ is differentiated in $\lambda^3$.
    \item Using the first derivatives, we construct Taylor expansions to first order in each of the spherical Stokes parameters.  These can be expressed
    \begin{align}
        I(\nu) &\approx I(\nu_0) + \frac{dI}{d\nu}(\nu-\nu_0), \\ 
        N(\nu) &\approx N(\nu_0) + \frac{dN}{d\nu}(\nu-\nu_0), \\ 
        \phi(\nu) &\approx \phi(\nu_0) + \frac{d\phi}{d\lambda^2}(\lambda^2-\lambda_0^2) \approx \phi(\nu_0) - \frac{2c^2}{\nu_0^3}\frac{d\phi}{d\lambda^2}(\nu-\nu_0),  \\
        \psi(\nu) &\approx \psi(\nu_0) + \frac{d\psi}{d\lambda^3}(\lambda^3-\lambda_0^3) \approx \psi(\nu_0) - \frac{3c^3}{\nu_0^4}\frac{d\psi}{d\lambda^3}(\nu-\nu_0). 
    \end{align}
    \item In order to approximate bandwidth-integration, we integrate analytically the equations for the Stokes parameters $\{I, Q, U, V\}$.  For a band spanning the frequencies $\nu_1$ and $\nu_2$, the bandwidth-averaged Stokes parameters are given by
    \begin{align}
        I_\mathrm{BW} &= \frac{1}{\nu_2-\nu_1}\int_{\nu_1}^{\nu_2} I(\nu) d\nu, \\ 
        Q_\mathrm{BW} &= \frac{1}{\nu_2-\nu_1}\int_{\nu_1}^{\nu_2} N(\nu)\cos[\phi(\nu)]\sin[\psi(\nu)] d\nu, \\
        U_\mathrm{BW} &= \frac{1}{\nu_2-\nu_1}\int_{\nu_1}^{\nu_2} N(\nu)\sin[\phi(\nu)]\sin[\psi(\nu)] d\nu, \\
        V_\mathrm{BW} &= \frac{1}{\nu_2-\nu_1}\int_{\nu_1}^{\nu_2} N(\nu)\cos[\psi(\nu)] d\nu. 
    \end{align}
    Analytic solutions to these integrals are provided in Appendix \ref{sec:analyticIntegrals}.
    \item Finally, the front and back emission halves are summed to produce the complete image.
\end{itemize}

\noindent In practice, $d\phi/d\lambda^2$, which encapsulates Faraday rotation, is the most important frequency-dependent effect to take into account.  Faraday conversion, which is encapsulated in $d\psi/d\lambda^3$, is not as strong as Faraday rotation in these models.  In some of our models, our split into the front and back halves is necessary for capturing the Faraday depolarization of back half of the emission (due to the Faraday thick mid-plane) while preserving the polarization of the front half.  As we will later explore in Figure \ref{fig:rm_distribution_functions}, some pixels may simultaneously have emission from the forward-jet with an RM of $\sim 10^3 \ \mathrm{rad}\;\mathrm{m}^{-2}$, and emission from the counter-jet with an RM of $\sim 10^9 \ \mathrm{rad}\;\mathrm{m}^{-2}$.  Consequently, $\phi(\lambda^2)$ for the pixel is highly non-linear, but can be approximated as the sum of two emitting regions with distinct RMs.

In some pixels, especially within the photon ring, we compute unreasonably high values of $|dN/d\nu|$ within our Taylor expansion.  This is due to multiple emission components in the same pixel Faraday rotating at different rates, which can periodically increase and decrease the total linearly polarized intensity in the pixel depending on the relative phases of these components.  We use $\frac{d^2N}{d\nu^2}$ to help identify such problematic pixels.  Let $\Delta\nu=\nu-\nu_0$ be the difference in frequency space between some frequency $\nu$ at which Stokes parameters are being evaluated and the frequency at which derivatives have been computed, $\nu_0 = 228 \ \mathrm{GHz}$.  Treating $\frac{1}{2}\frac{d^2N}{d\nu^2} \Delta\nu^2$ as an upper limit on the error of $N(\nu)$, we freeze the value of $N$ in pixels where $|\frac{dN}{d\nu} \Delta\nu| < |\frac{1}{2}\frac{d^2N}{d\nu^2} \Delta\nu^2|$ and $|\frac{d\ln(N)}{d\nu}| > |\frac{1}{\Delta\nu}|$.  That is, we freeze $N$ if its first derivative implies that it would grow or shrink by more than a factor of $e$ across $\Delta\nu$ and the absolute error on the change of $N$ may be larger than the change of $N$ itself.  We apply an identical condition on the derivative of $I$, which almost always affects pixels also affected by the condition on Stokes $N$.

We examine to what extent this criterion is applied in our final snapshot images, and find that it affects only a minority of pixels.  Among the pixels which amount to 99 per cent of the image integrated $I$, we find that at most 5 per cent of the pixels are affected by this correction, typically in the photon ring.  This largest fraction occurs in the SANE, $a=-0.94$, $R_\mathrm{high}=80$ model, which as we shall show contains the emission with the largest Faraday depths.  In some of the other models, none of the pixels are affected by this criterion.

In Figure \ref{fig:taylor_expansion_example}, we demonstrate the efficacy of our method by plotting the linearly polarized intensity ($\sqrt{Q^2+U^2}$) for the SANE, $a=+0.94$, $\mathrm{R}_\mathrm{high}=160$ model.  The top left panel shows the result of a single-frequency calculation at 228 GHz, while the top right image shows the result of averaging 255 images across a 4 GHZ bandwidth between 226 and 230 GHz.  In the bandwidth-averaged image, much of the large-scale emission has been suppressed and two streaks of emission in the upper right of the image have become more prominent.  The total spatially unresolved linear polarization fraction has dropped by a factor of 2.4.  Although the counter-jet dominates the linearly polarized emission in the single-frequency image due to lensing \citep{Dexter+2012}, most of that emission is scrambled away due to the large Faraday depth in the mid-plane \citep{Moscibrodzka+2017}.

The bottom left panel is the result of our Taylor approximated model based on 6 images (3 frequencies, 2 sides), which successfully captures the depolarization of the counter-jet, but not the forward-jet.  In the bottom right panel, we subtract this model from the averaged images and plot the residual.  Weighting by the linear polarized intensity of the properly averaged image, total linear polarization is recovered on average with an absolute error of $0.32 \ \mu\mathrm{Jy}$ and a relative error of $0.081$.  In contrast, the single-frequency image has an average absolute error $5.7 \ \mu$Jy and relative error of $4.0$.  In this case, the single-frequency image also over-estimates the total linear polarization somewhat, even if the source is spatially unresolved.  When Stokes parameters are summed across the entire image, the properly averaged image has a linear polarization fraction of $8.3 \times 10^{-3}$, the Taylor approximated image has a linear polarization fraction of $7.3 \times 10^{-3}$, and the single-frequency image has a linear polarization fraction of $1.9 \times 10^{-2}$.  A model containing at least two separate zones is necessary in order to reproduce the depolarization of the counter-jet without also depolarizing the forward-jet.  The remaining residuals require more than two regions to fully capture the complexity of the Faraday rotating structure along the line-of-sight.  Notice that the most significant errors occur within the photon ring or its interior, where geodesics pass through the mid-plane multiple times.

\begin{figure*}
  \centering
  \includegraphics[width=\textwidth]{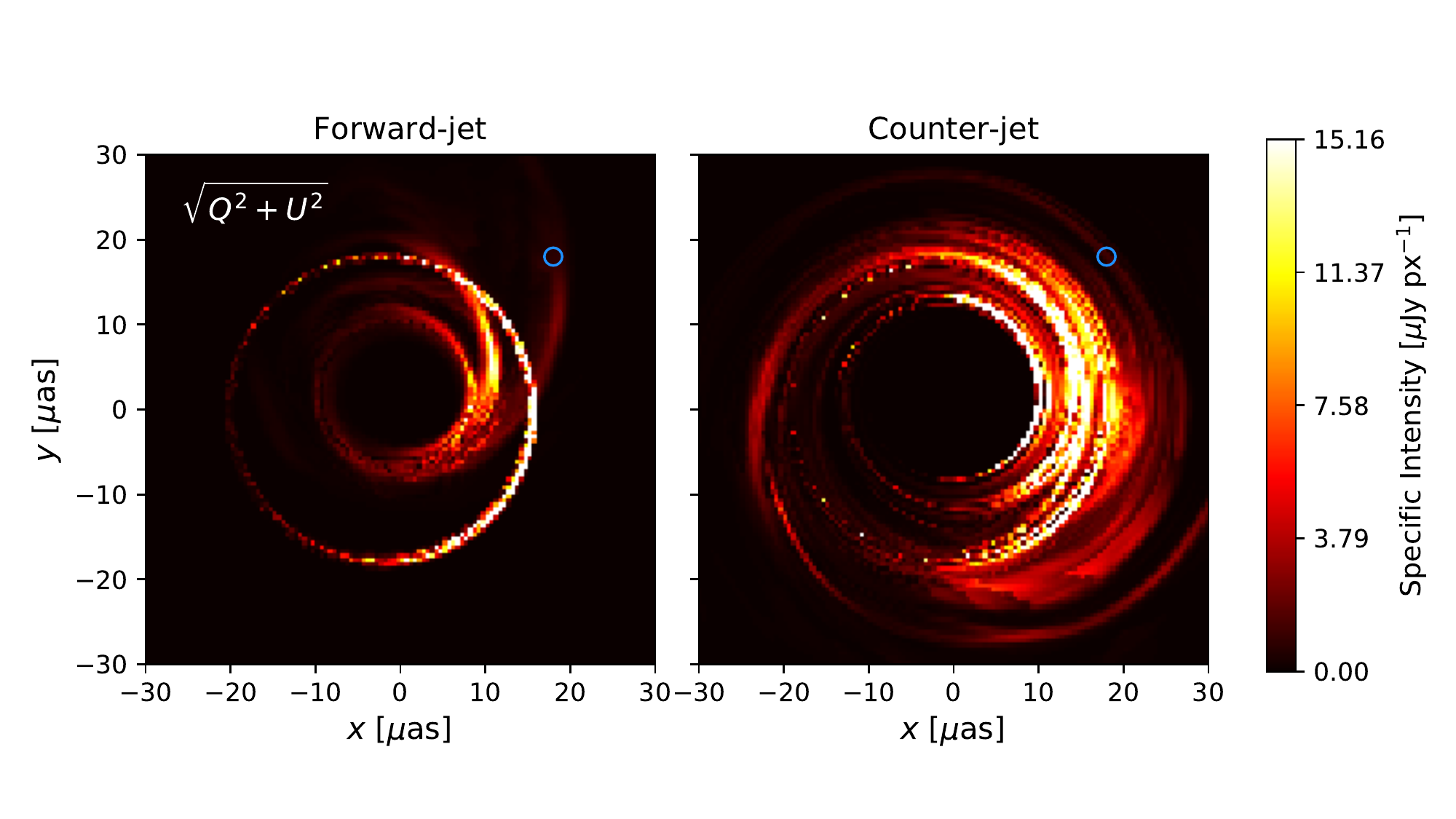}
  \includegraphics[width=\textwidth]{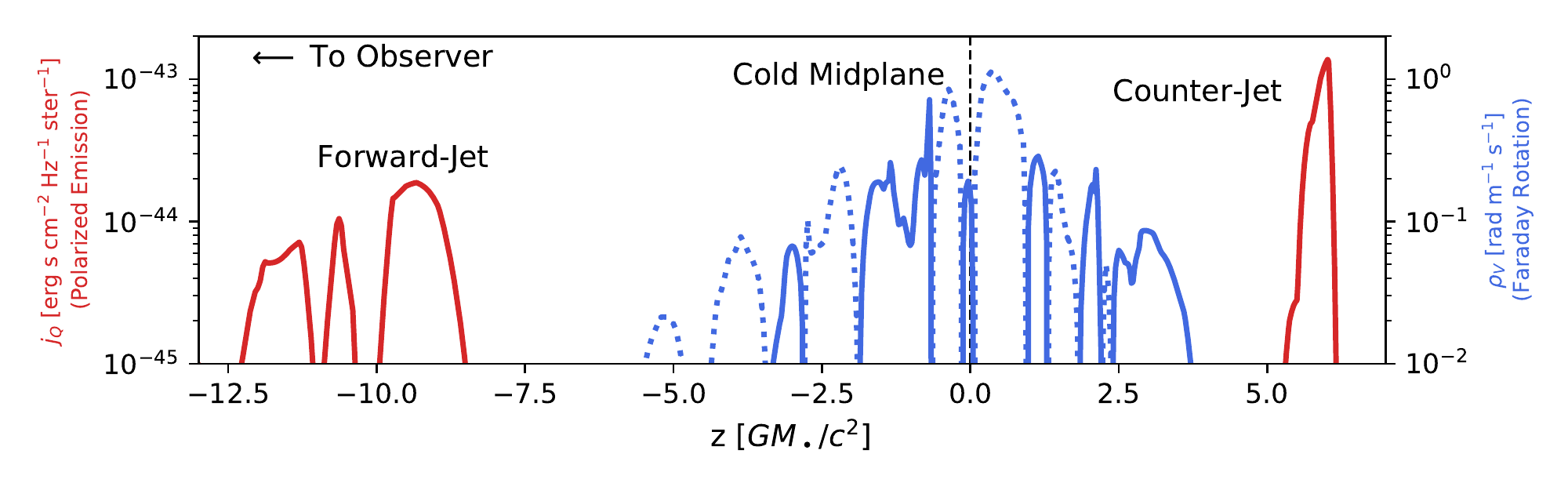}
  \caption{Linearly polarized intensity ($\sqrt{Q^2+U^2}$) for the SANE, $a=+0.94$, $\mathrm{R}_\mathrm{high}=160$ model at 227 GHz (0 bandwidth), broken into forward-jet (Top Left) and counter-jet (Top Right) emission components, which appear very different due to lensing.  For the geodesic marked by a blue circle, we plot in the lower panel the frame-invariant versions of the radiative transfer coefficients that correspond to polarized emissivity $(j_Q \rm{, red})$ and Faraday rotation $(\rho_V \rm{, blue})$.  For $\rho_V$, dotted lines correspond to negative values, and there are many sign flips due to the turbulent nature of a SANE disc.  Emission from the counter-jet passes through the cold mid-plane on the way to the observer, which produces a very large RM.  Thus, when properly averaging over bandwidth, the counter-jet emission is suppressed by bandwidth depolarization.  (See Fig. \ref{fig:taylor_expansion_example}, top right panel.)   \label{fig:taylor_expansion_example_breakdown}}
\end{figure*}

In Figure \ref{fig:taylor_expansion_example_breakdown}, we further decompose the single-frequency 228 GHz image into emission from its forward-jet and from counter-jet components.  In the lower panel, we examine frame-invariant\footnote{Since $\nu$ varies along the geodesic, it is useful to define the frame-invariant quantities $\rho_V = \nu\rho_{\nu,V}$ and $j_Q = j_{\nu,Q}/\nu^2$ \citep[for additional discussion, see][]{Moscibrodzka&Gammie2018}.} radiative transfer coefficients in the pixel marked by a blue circle in these images.  In this pixel, polarized emission is emitted roughly equally by forward-jet and counter-jet components.  However, the counter-jet emission must pass through the enormous Faraday depth of the cold mid-plane, with $\mathrm{RM} > 10^9 \ \mathrm{rad} \; \mathrm{m}^{-2}$ in some regions.  The polarized intensity of emission passing through material with a rotation measure of RM and a bandwidth of $\Delta\nu$ is suppressed by a factor $f_\mathrm{BW}$ given by

\begin{equation}
    f_\mathrm{BW} = \sinc\left(\frac{2c^2}{\nu_0^3}\mathrm{RM}\Delta\nu\right),
    \label{eqn:bw_depolarization}
\end{equation}

\noindent where $\sinc(x) = \sin(x)/x$, $c$ is the speed of light, and $\nu_0$ is the central frequency, assuming narrow fractional bandwidth and uniform sampling in $\nu$.  We define the critical rotation measure $\mathrm{RM}_\mathrm{crit}$ via

\begin{equation}
    \frac{1}{2} = \sinc\left(\frac{2c^2}{\nu_0^3}\;\mathrm{RM}_\mathrm{crit}\;\mathrm{BW}\right).
    \label{eqn:rm_crit}
\end{equation}

\noindent That is, $\mathrm{RM}_\mathrm{crit}$ is the minimum $\mathrm{RM}$ required to suppress linear polarization by a factor of 2.  So far, EHT observations have been performed using a central frequency of 228 GHz and a bandwidth of 4 GHz.  Using these EHT values, we find that $\mathrm{RM}_\mathrm{crit} = 3 \times 10^7 \ \mathrm{rad} \; \mathrm{m}^{-2}$.  Since $10^9 \ \mathrm{rad} \; \mathrm{m}^{-2} \gg \mathrm{RM}_\mathrm{crit}$, the counter-jet's polarization is significantly suppressed.  Notice that the photon ring emission from the forward-jet is also suppressed, since this emission must also pass through the cold mid-plane.  \citet{Jimenez-Rosales&Dexter2018} determine that strong Faraday effects also scramble the image on a pixel-by-pixel basis, resulting in beam depolarization.  This spatial decoherence helps compensate for bandwidth depolarization when blurred images are constructed.

\subsection{Rotation Measure}
\label{sec:rotation_measure}

In \S\ref{sec:taylor_expansion}, we compute Stokes parameters and their derivatives at a central frequency of 228 GHz and a small bandwidth of 2 MHz.  From these, we can directly compute the RM in each pixel at 228 GHz via

\begin{equation}
    \mathrm{RM} \equiv \frac{d\chi}{d\lambda^2} = \frac{d}{d\lambda^2}\frac{1}{2}\arctan\left(\frac{U}{Q}\right) = \frac{1}{2}\frac{U'Q-Q'U}{Q^2+U^2},
    \label{eqn:rm_calculation}
\end{equation}

\noindent which follows from a straightforward application of the chain rule; here the $'$ symbols denote $d/d\lambda^2$.  When we discuss the RM of individual pixels, this is how RM is computed.  Using a small band of 2 MHz, the maximum measurable RM in an individual pixel is $|\mathrm{RM}|_\mathrm{max} = \pi/\Delta\lambda^2 \approx \pi \nu_0^3/(2c^2\Delta\nu) = 1.0\times10^{11} \ \mathrm{rad} \; \mathrm{m}^{-2}$. 

However, when discussing the RM for spatially unresolved measurements in subsequent sections, it is important to recognise the complicated evolution of $\chi(\lambda^2)$ that results from the complex RM structure we uncover in \S\ref{sec:spatial_variation}.  When assigning a single value of the RM to an entire spatially unresolved image, we use the Taylor expansion methodology developed in \S\ref{sec:taylor_expansion} to approximate 16 polarized images, each integrated over a bandwidth of $0.25$ GHz, equally spaced in $\lambda^2$ space between 226 and 230 GHz to emulate EHT observations.  These images are used to compute $\chi(\lambda^2)$, and $\Delta\chi$ is computed from the endpoints of the band.  We correct for phase wraps by adding or subtracting $\pi$ to $\chi(\lambda^2)$ as necessary to obtain the correct sign of $d\chi/d\lambda^2$ based on Equation \ref{eqn:rm_calculation}.  Using 16 bands of width 0.25 GHz, the maximum measurable RM is $|\mathrm{RM}|_\mathrm{max} = \pi/\Delta\lambda^2 \approx \pi \nu_0^3/(2c^2\Delta\nu) = 8.3\times10^8 \ \mathrm{rad} \; \mathrm{m}^{-2}$. 

\section{Results}
\label{sec:results}

\subsection{Structure of Spatially Resolved RM}
\label{sec:spatial_variation}

In this section, we study the RM of each of our models within single simulation snapshots.  We find significant spatial variation across the image due to inhomogeneities in the accretion flow on event horizon scales.  At different locations, RM can vary by orders of magnitude and even flip sign.  As a result, for spatially unresolved polarized measurements, some of these models may exhibit highly non-linear $\chi(\lambda^2)$, making them difficult to characterise with a single RM.

\begin{figure*}
    \centering
    \includegraphics[width=\textwidth]{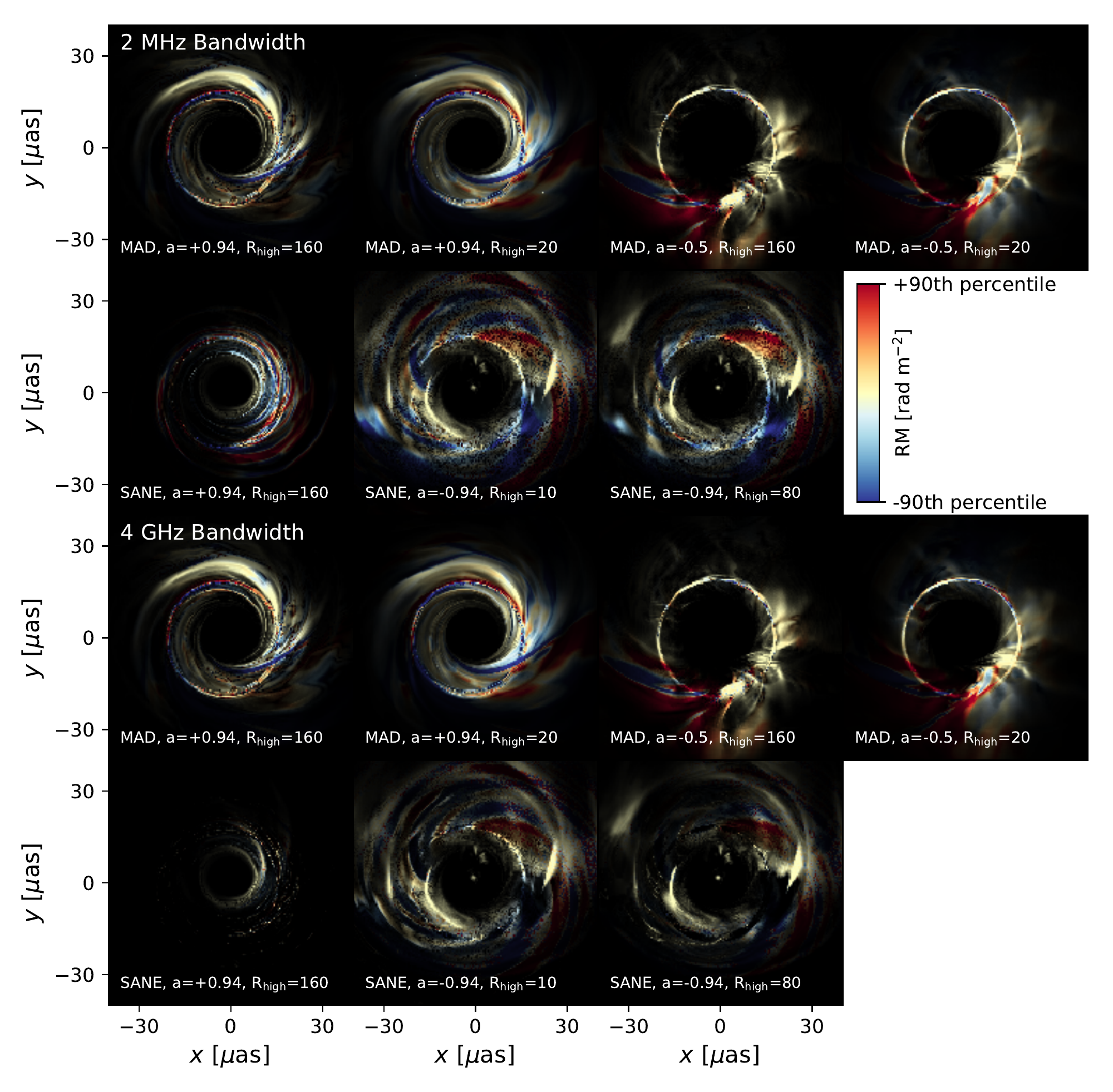}
    \caption{Visualisation of the RM structure for the snapshots plotted in Figure \ref{fig:intensity_examples}.  The brightness of each pixel is proportional to the linearly polarized intensity in the pixel (again saturating 0.3 per cent of the pixels), while the coloration encodes the RM.  The colour scale is normalised separately for each model, spanning $\pm$ the 90th-percentile of $|\mathrm{RM}|$ of the pixels which have at least 50 per cent of the maximum linear polarized intensity.  Red regions have positive RM, while blue regions have negative RM.  The top two rows are not bandwidth corrected, while the bottom two are.  This can dramatically affect the SANE models, but has little effect on the MAD models.  Complex structure and sign flips are a generic prediction of these simulations.
    \label{fig:rm_examples}}
\end{figure*}

In Figure \ref{fig:rm_examples}, we visualise the spatially resolved RM structure of the images presented in Figure \ref{fig:intensity_examples}.  In this figure, the brightness of each pixel scales with the linearly polarized specific intensity $(\sqrt{Q^2+U^2})$, while the coloration encodes the RM.  As in Figure \ref{fig:intensity_examples}, 0.3 per cent of the pixels are saturated to more clearly display low surface brightness structures. The colour-scale is normalised separately for each model, spanning $\pm$ the 90th-percentile of $|\mathrm{RM}|$ of the pixels which have at least 50 per cent of the maximum linear polarized intensity plotted.  Red regions have positive RM, while blue regions have negative RM.  The top two rows depict the RM with $\Delta \nu = 2$ MHz, where our Taylor expansion model is constructed.  The bottom two rows plot the RM across a 4 GHz band, a more realistic bandwidth, within which bandwidth depolarization becomes important for some models.  4 GHz images are plotted with the same brightness and RM scale as their 2 MHz counterparts.

\begin{figure}
  \centering
  \includegraphics[width=0.5\textwidth]{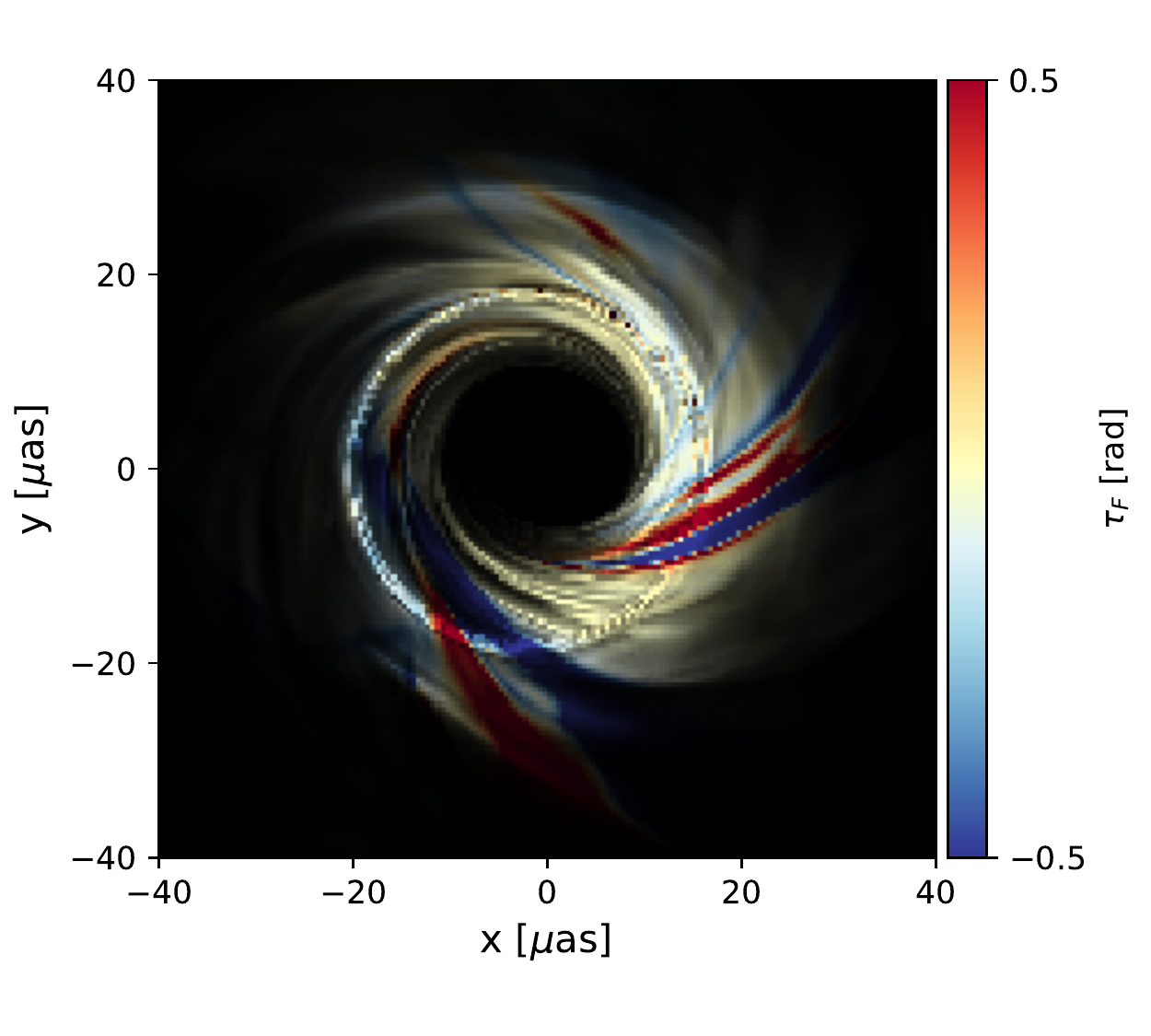}
  \caption{Intensity-weighted Faraday depth in one snapshot of the MAD, $a=0.94$, $R_\mathrm{high}=20$ model, revealing clear sign flips in the magnetic field parallel to the line-of-sight.  The colour of each pixel encodes the total Faraday depth, while the brightness is proportional to the linear polarized intensity.  The sign of the Faraday depth directly encodes the direction of the magnetic field parallel to the geodesic.  Subsequent 3D visualisation of this snapshot reveals that these sign flips occur in the tangential magnetic field in the plane of the disc.  \label{fig:sign_flips}}
\end{figure}

As shown in this figure, complex spatial variation and frequent sign-flips are a generic feature of these RM maps.  This behaviour is not surprising in SANE models, which are characterised by weaker, disordered magnetic fields, but is less expected in MAD models, which are characterised by strong poloidal fields.  In one suggestive snapshot, we confirm that these RM sign-flips are due to sign-flips in the magnetic field with respect to the geodesic.  Figure \ref{fig:sign_flips} plots the intensity-weighted Faraday depth in each pixel, $\tau_F = \int \rho_V ds$, for a snapshot of the MAD, $a=0.94$, $R_\mathrm{high}=20$ model.  Here, $\rho_V$ is the (frame-invariant) radiative transfer coefficient responsible for Faraday rotation, and $s$ is the affine parameter describing the geodesic.  The sign of this quantity, shown to exhibit both positive to negative values, directly encodes the direction of the magnetic field with respect to the photon trajectory.  RM sign flips have been predicted by earlier MHD simulations without GR, but only at large inclinations \citep{Sharma+2007}.

By performing a 3D visualisation of this snapshot with {\tt VisIt} \citep{VisIt2012}, we find that sign flips in the magnetic field occur in its tangential and radial components when crossing the disc mid-plane.  Near the event horizon, field lines on the northern hemisphere point west, while field lines on the southern hemisphere point east, although they point north overall (in the positive z-direction).  This is a natural consequence of the tangential stretching of vertical field lines as they are dragged into the BH by accreting material, as well as frame dragging \citep[see e.g.,][for helpful schematics]{Contopoulos+2009,Gabuzda2018}.  Since this sign flip occurs when crossing the mid-plane, accreting streams of gas that straddle the mid-plane can exhibit streaks of positive RM adjacent to streaks of negative RM.

Returning to Figure \ref{fig:rm_examples} and comparing the 2 MHz bandwidth visualisations to the 4 GHz bandwidth visualisations, SANE models are more strongly affected by bandwidth depolarization than MAD models.  In each of the SANE models, counter-jet polarized emission is especially suppressed.  This more strongly changes the morphology of the prograde SANE than the retrograde SANEs, because the image morphologies of the counter-jet and forward-jet in the retrograde SANEs are more similar.  

Closely examining the 2 MHz counter-rotating SANE models, one may notice that the linearly polarized intensity appears to vary strongly among adjacent pixels, causing the appearance of ``static'' across the image.  This effect is an artefact of single-frequency radiative transfer calculations, an instance of Faraday rotation randomizing not only the phase of the linear polarization, but also its amplitude.  In this model, the linear polarization in each pixel is well-approximated by the sum of its forward-jet and counter-jet components, neither of which exhibit this ``static'' if plotted individually.  The phase of the counter-jet emission is effectively randomized by the enormous Faraday depth in the mid-plane.  Depending on the relative phase, the rotated counter-jet polarized emission may sometimes cancel with the forward-jet polarized emission.  This effect would not occur in real observations integrated over a finite bandwidth.

\subsection{RM Distribution Functions for M87*}
\label{sec:distribution_functions_17}

\begin{figure*}
  \centering
  \includegraphics[width=\textwidth]{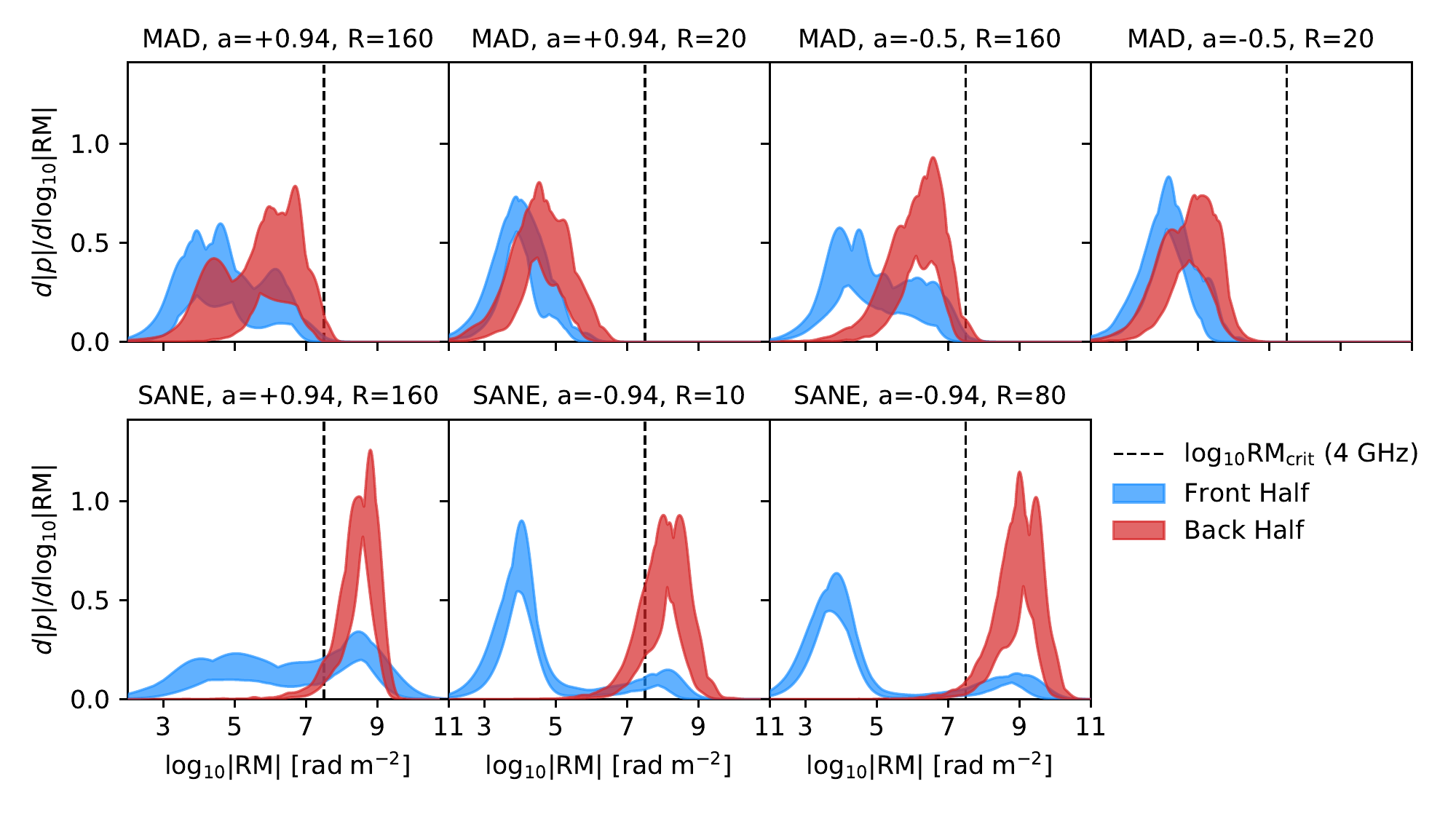}
   \caption{Distribution functions of $\log_{10} |\mathrm{RM}|$ for our models, weighted by the linear polarized intensity of each pixel.  Emission originating from front side of the emitting region is coloured blue, while emission originating from the back side is coloured red.  The relatively cold mid-plane is the dominant source of Faraday rotation in these models, which can result in significant offsets in these distributions between the two halves, especially in SANE models.  Forward-jet emission in SANE models exhibits a second peak at high $\log_{10} |\mathrm{RM}|$ due to photon ring geodesics, which do pass through the mid-plane. \label{fig:rm_distribution_functions}}
\end{figure*}

In these models, polarized emission exhibits a wide range of Faraday depths, and the front and back halves of the emitting regions can differ by many orders of magnitude.  In Figure \ref{fig:rm_distribution_functions}, we plot the distribution functions of $\log_{10}|\mathrm{RM}|$ among the pixels of each model during their final snapshot, weighted by each pixel's linear polarized intensity.  This quantity, which we denote as $d|p|/d\log_{10}|\mathrm{RM}|$, is closely related to $F(\phi)$ in rotation measure synthesis theory, the complex polarized surface brightness per unit Faraday depth \citep{Burn1966,Brentjens&deBruyn2005}.  Unlike $F(\phi)$, we take absolute values, normalise with respect to the total emitted polarized surface brightness, and adopt logarithmically spaced bins.

In each pixel, $|\mathrm{RM}|$ is computed directly from the gradient at 228 GHz across $\Delta\nu = 2 \ \mathrm{MHz}$. These distributions are computed for each of the 11 snapshots that we study, and the filled regions span the range permitted by all of these snapshots.  Blue regions only include emission originating from the front half of the emitting region, while red regions only include emission originating from the back half.  The dashed vertical line marks $\mathrm{RM}_\mathrm{crit}$ (eqn. \ref{eqn:rm_crit}) for a 4 GHz bandwidth.  Any emission to the right of this line is significantly bandwidth depolarized.  The distributions of front half of the emission region can be significantly displaced from that of the back half, especially in SANE models.  Again, this is due to the large Faraday depth of the sub-relativistic mid-plane in these models.  In some models, including all of the SANEs, the forward-jet distributions exhibit two distinct peaks.  The peak at higher $|\mathrm{RM}|$ is due to photon ring orbits, which pass through this mid-plane.  Note the extreme difference in $|\mathrm{RM}|$ between front and back components in the jet-dominated retrograde SANE models.  These models may exhibit much lower spatially unresolved $|\mathrm{RM}|$ than would be expected by integrating their Faraday depths across geodesics, since this Faraday depth mainly only affects the counter-jet.

\subsection{Non-linear Structure of Spatially Unresolved RM}

\begin{figure*}
  \centering
  \includegraphics[width=\textwidth]{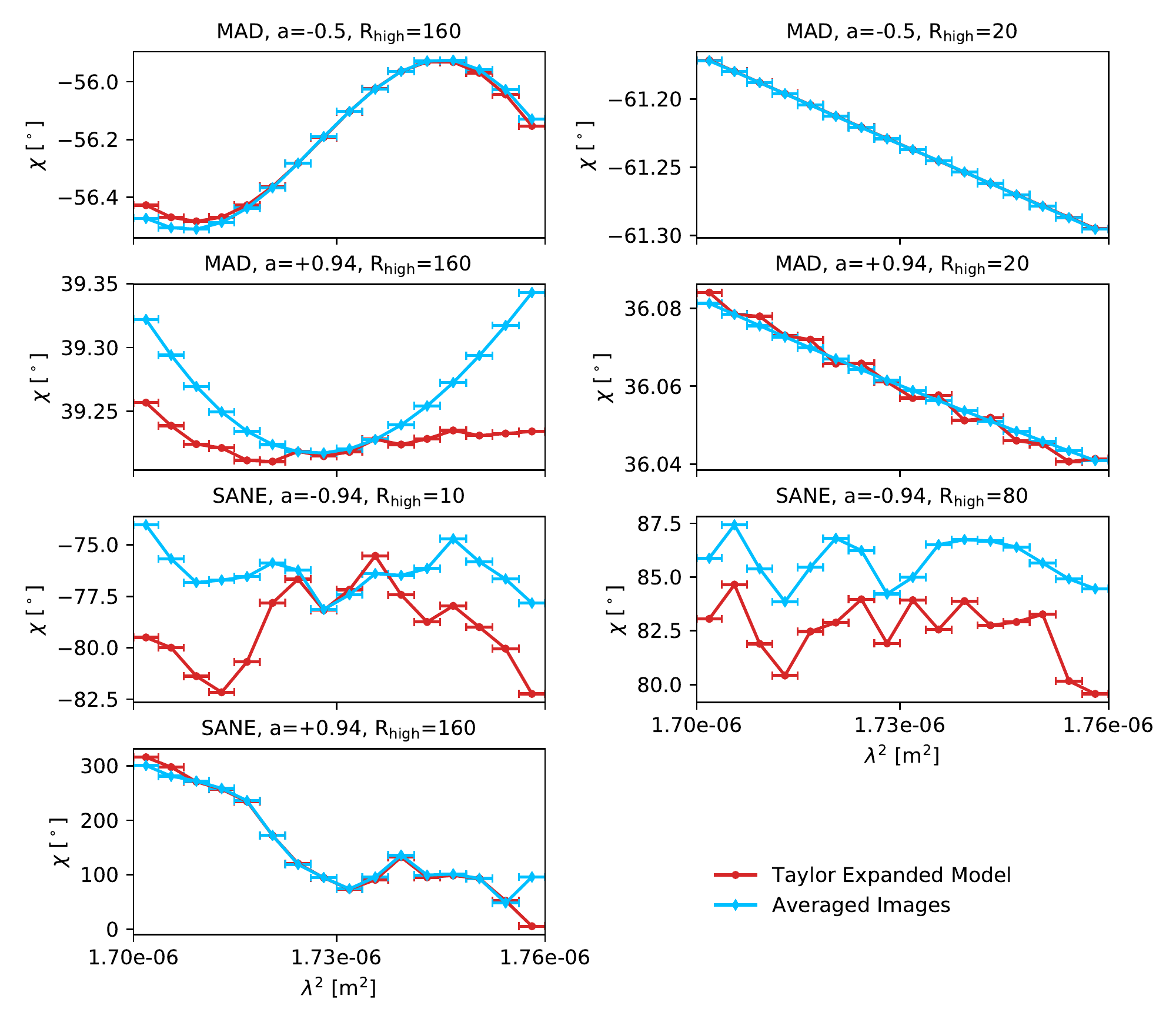}
   \caption{Non-$\lambda^2$ behaviour of the 7 models we consider during their final simulation snapshot.  Within 16 bands that are each 0.25 GHz wide, we plot $\chi(\lambda^2)$ of a band-averaged image.  In blue, we plot the result from 255 images evenly spaced between 226 and 230 GHz.  Within each band, the appropriate subset of images is averaged.  In red, we plot the result from our Taylor expansion model based on 6 images around 228 GHz.  Most of these models exhibit non-linear behaviour even within this narrow fractional bandwidth.   \label{fig:nonlambda2_16bands}}
\end{figure*}

In \S\ref{sec:spatial_variation}, we demonstrate that GRMHD models exhibit rich spatial structure, whereby the RM can vary by many orders of magnitude and flip sign.  As a consequence, $\chi(\lambda^2)$ rotates at very different rates at different locations within the image, which may result in clear departures from a $\lambda^2$ law if these structures are not spatially resolved.  In Figure \ref{fig:nonlambda2_16bands}, we investigate this non-linearity by plotting $\chi(\lambda^2)$ for the final snapshots of our 7 models.  In blue, we laboriously compute 255 individual images across the 4 GHz bandwidth, then average Stokes parameters within 16 smaller 0.25 GHz bands to estimate $\chi(\lambda^2)$.  In red, we instead use the Taylor expansion model based on 6 images and analytic integrals described in \S\ref{sec:taylor_expansion} to estimate $\chi(\lambda^2)$.

All models exhibit significantly non-$\lambda^2$ behaviour even within this small fractional bandwidth except for the MAD $\mathrm{R}_\mathrm{high}=20$ models.  SANE models are especially non-linear, and in fact exhibit spectrally unresolved structure in $\chi(\lambda^2)$ even among the 255 images separated in frequency by 16 MHz.  This is because as shown in Figure \ref{fig:rm_distribution_functions}, a significant amount of the intensity has an associated $|\mathrm{RM}|$ of $\approx 10^9 \ \mathrm{rad}\;\mathrm{m}^{-2}$.  The MAD $\mathrm{R_\mathrm{high}} = 160$ models exhibit relatively mild non-linearity, since $|\mathrm{RM}|$ only just approaches $|\mathrm{RM}_\mathrm{crit}|$ in Figure \ref{fig:rm_distribution_functions}.

Recall that we use $\Delta\chi/\Delta\lambda^2$ across the bandwidth from our Taylor expansion model to assign spatially unresolved RMs to images.  This figure reveals some of this model's limitations.  The model poorly reproduces $\chi(\lambda^2)$ for the MAD, $a=+0.94$, $\mathrm{R}_\mathrm{high}=160$ model, possibly due to 228 GHz being a local extremum of $\chi(\lambda^2)$ where the first derivative is small.  Interestingly, the retrograde SANE Taylor expansion models appears to broadly follow the structure of the true $\chi(\lambda^2)$, but with a vertical offset.  This is due to incorrect evolution of emission superposed on top of the photon ring.  Our Taylor expansion model assigns a large $d\phi/d\lambda^2$ to photon ring pixels, but these pixels also contain forward-jet emission that does not pass through the mid-plane.  In the SANE, $a=-0.94$, $\mathrm{R}_\mathrm{high}=80$ model, this superposed component is immediately bandwidth depolarized and subtracted, leading to the offset.  This effect is more delayed as a function of $\Delta\lambda^2$ in the SANE, $a=-0.94$, $\mathrm{R}_\mathrm{high}=10$ model, which by construction has a warmer mid-plane and therefore less Faraday rotation.  Fortunately, this effect is symmetric about the Taylor expansion point at 228 GHz and we can still recover the spatially unresolved RM from $\chi$ at the end points of the band.

\subsection{RM as a Measure of Accretion Rate}
\label{sec:rm_mdot}

RM is often used to approximate the accretion rate $\dot{M}_\bullet$, based on simple analytic models \citep{Marrone+2006}.  These models are based on advection or convection dominated accretion flows  \citep{Narayan&Yi1994,Narayan+2000,Quataert&Gruzinov2000a} and make many simplifying assumptions.  These include spherical symmetry, equipartition of energy, and rather arbitrary inner and outer radii to truncate the model.  Adapted from \citet{Marrone+2006}, 

\begin{align}
    \begin{split}
    \mathrm{RM} &= \left( 3.4\times10^{19} \ \mathrm{rad} \; \mathrm{m}^{-2} \right) \left[ 1-(r_\mathrm{out}/r_\mathrm{in})^{-(3\beta-1)/2}) \right] \\
    &\times \left(\frac{M_\bullet}{3.5\times 10^6 \ M_\odot} \right)^{-2}\left(\frac{2}{3\beta-1}\right) r_\mathrm{in}^{-7/4} \dot{M}_\bullet^{3/2},
    \end{split}
\end{align}

\noindent where we have corrected the exponent of $r_\mathrm{in}$, as noted by \citet{Macquart+2006}.  Here, $r_\mathrm{out}$ and $r_\mathrm{in}$ are radii used to truncate the model in units of Schwarzschild radii, and $\beta\in [0.5,1.5]$ describes the slope of the density profile.  Note that the model is insensitive to the choice of $r_\mathrm{out}$ unless $r_\mathrm{out}/r_\mathrm{in} \approx 1$.  $\dot{M}_\bullet$ carries units of $M_\odot \; \mathrm{yr}^{-1}$.

\begin{figure*}
  \centering
  \includegraphics[width=\textwidth]{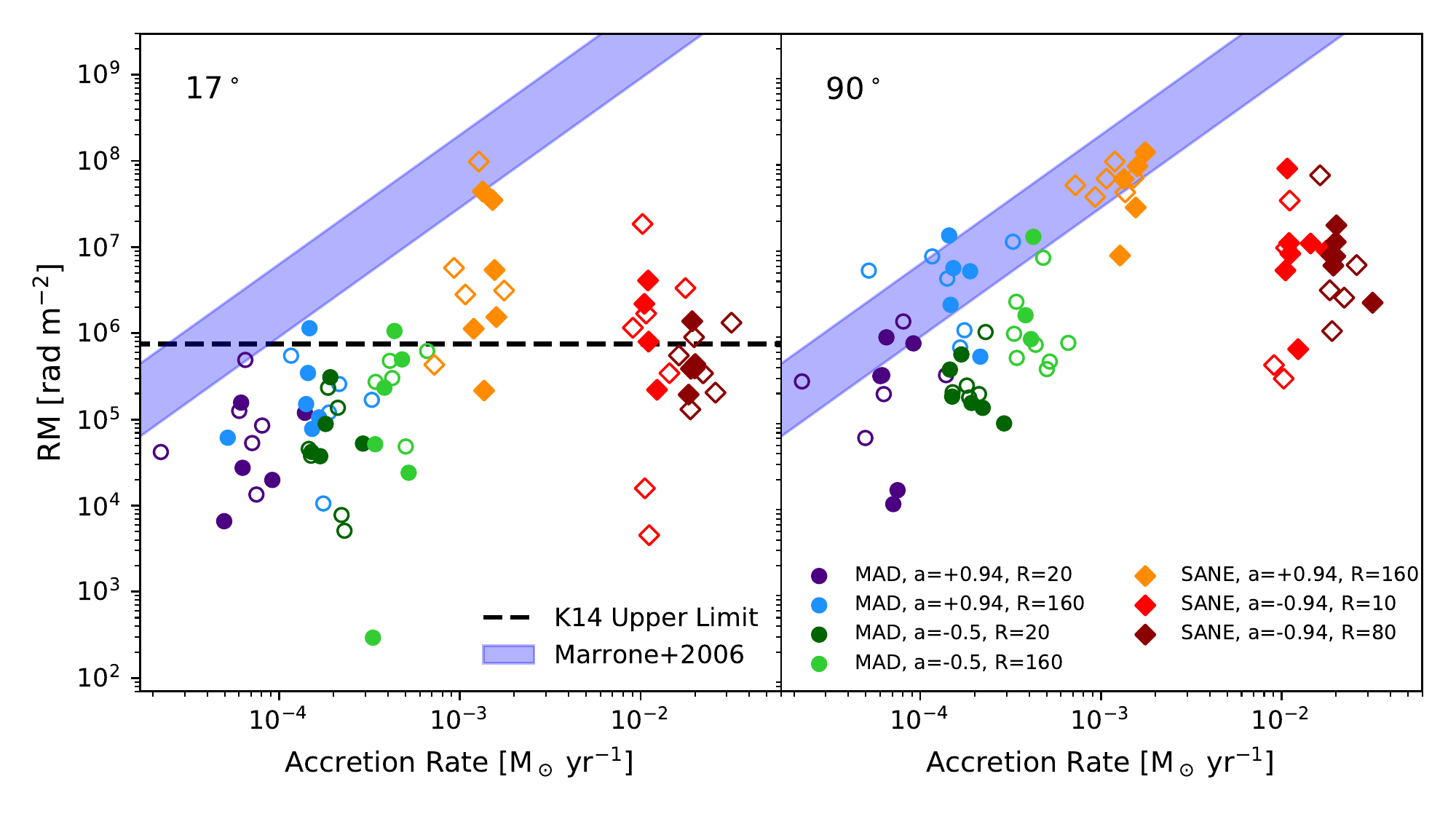}
  \caption{RM as a function of accretion rate for the 7 models considered in this paper.  Filled symbols have positive RM, while open symbols have negative RM.  In the left panel, the observer is oriented at a 17$^\circ$ inclination, while in the right panel, the observer is oriented at 90$^\circ$.  At an inclination of $17^\circ$, we find that accretion rates are systematically higher than those that would be inferred by simple analytic models \citep{Marrone+2006}.  At 90$^\circ$, we find RMs in better agreement with analytic models, although there remains substantial scatter.  Retrograde SANE models are outliers, since their forward-jet emission does not intercept the large Faraday depth in the mid-plane.
  \label{fig:rm_mdot_17}}
\end{figure*}

Here, we test the relationship between RM and accretion rate in our suite of images for M87*.  In Figure \ref{fig:rm_mdot_17}, we plot spatially unresolved RM versus accretion rate for the 7 models considered in this work.  Each symbol of the same colour represents a different snapshot of the same model.  Positive RMs are plotted with filled symbols, while negative RMs are plotted with open symbols.  Notice the ubiquitous flips in the sign of the RM that occur in all models.  The filled blue region demarcates the relation in \citet{Marrone+2006}, spanning variations of the slope of the density profile, $\beta\in[0.5,1.5]$.  We set $r_\mathrm{in}=3$ and $r_\mathrm{out} = \infty$.  The dashed line shows the upper limit from \citet{Kuo+2014}.  A $17^\circ$ inclination appropriate for M87* is shown on the left, while for comparison a $90^\circ$ inclination is shown on the right.

Overall, we find that a spatially unresolved RM is a poor predictor of the accretion rate, especially if the correct model is not known a priori.  RM and the accretion rate differ by orders of magnitude both within and among the different models.  Even within a single model, there is no correlation between RM and accretion rate, which we explore in more detail for one model in \S\ref{sec:case_study}.  Since their higher accretion rates imply higher number densities, SANE models typically have larger spatially unresolved RMs than MAD models.  However, these models exhibit such strong time variability that they cannot be distinguished solely by the upper limit of \citet{Kuo+2014}.  Rather, repeat observations on timescales of months to years will be necessary to characterise the distribution of $|\mathrm{RM}|$ over time and detect potential sign flips.

For a given accretion rate, the GRMHD models in this work produce much lower RM than the analytic model of \citet{Marrone+2006} at an inclination of $17^\circ$.  As we further explore in \S\ref{sec:inclination}, this is because these simulations are viewed through an evacuated funnel region at low inclination.  With an inclination of $90^\circ$, the $|\mathrm{RM}|$ more closely matches that predicted by \citet{Marrone+2006}, although they still remain systematically offset.  A similar inclination dependence is found for SANE models in \citet{Moscibrodzka+2017}.  The retrograde SANE models remain the most offset from the analytic model.  Even at $90^\circ$, the large Faraday depth occurs in an area with little emission, since the electrons are assigned low temperatures in the mid-plane.

\subsection{Dependence of RM on Inclination}
\label{sec:inclination}

\begin{figure*}
  \centering
  \includegraphics[width=\textwidth]{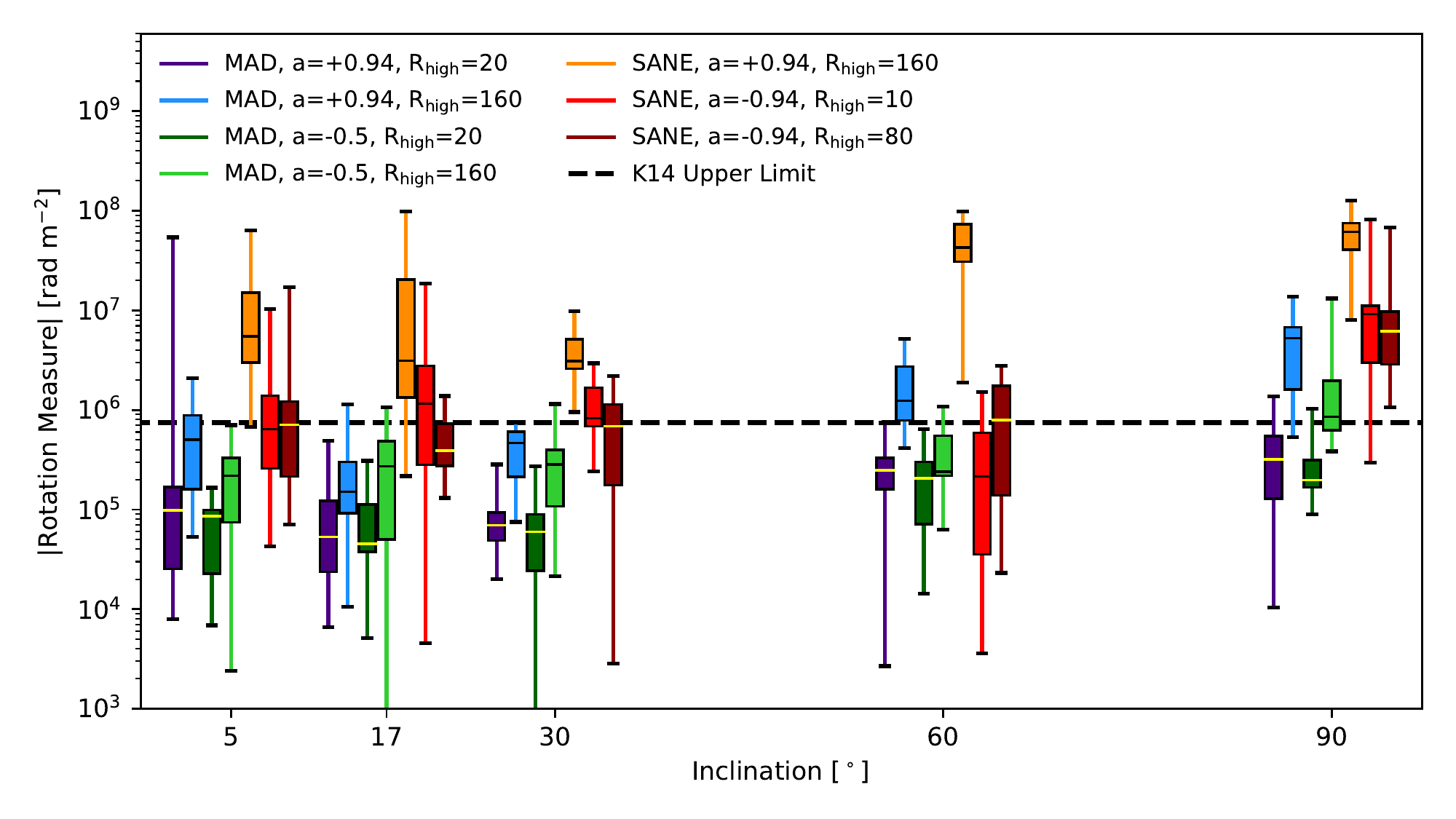}
  \caption{RM as a function of inclination.  11 snapshots are shown for each model.  For M87*, 17 degrees is considered the most likely inclination based on its large-scale jet.  We plot the \citet{Kuo+2014} upper limit on the RM with a dashed line.  Boxes contain the 25th to 75th quantiles, horizontal black or yellow lines mark the median, and the error bars span the full range of the 11 snapshots considered.  We report a noticeable inclination dependence, due to the evacuated jet region through which the BH is observed at low inclination.  \label{fig:rm_inclination}}
\end{figure*}

Here, we study the dependence of RM on inclination in greater detail.  In Figure \ref{fig:rm_inclination}, we plot the distributions of RM for all 11 snapshots of all 7 models at 5 different inclinations.  In this plot, boxes contain the 25th to 75th quantiles, the horizontal black or yellow line marks the median, and the error bars span the full range of the 11 snapshots studied.  This plot omits the sign flips observed in Figure \ref{fig:rm_mdot_17}, but we comment that they remain ubiquitous at all inclinations for these calculations which terminate at a radius of $20 \; GM_\bullet/c^2$.

As in \citet{Moscibrodzka+2017}, we find that the absolute value of the RM depends on the inclination angle, but it is not large compared to the substantial scatter between snapshots.  This weaker dependence is likely due to the small radius at which we truncate our calculations.  We notice no differences between our calculations at low inclination:  $5^\circ$, $17^\circ$, and $30^\circ$.  This is fortunate for our study of M87*, as this indicates that we need not be concerned with small deviations from our fiducial inclination of $17^\circ$.

At low inclinations, we view the accretion flow through an evacuated funnel region with less Faraday rotating material than at high inclinations.  We demonstrate this by calculating the characteristic distance of Faraday rotating material as a function of inclination.  By modifying the {\tt ipole} source code, we compute for each geodesic

\begin{equation}
    \langle R_{FR} \rangle \equiv \int R_{BL} |\rho_V| LP ds \bigg/ \int|\rho_V| LP ds,
\end{equation}

\noindent where $\rho_V$ is the frame-invariant radiative transfer coefficient responsible for Faraday rotation, $R_{BL}$ is the radius of the material in Boyer-Lindquist (or equivalently Kerr-Schild) coordinates, $LP = \sqrt{Q^2+U^2}$ is the total amount of linearly polarized emission that has been emitted along the geodesic so far (on the way to the camera), and $s$ is the affine parameter describing the geodesic.  In other words, this is the characteristic distance of Faraday rotating material, weighted by the fraction of the final linearly polarized emission that has already been added to the pixel on the way to the camera.  Once $R_{FR}$ is computed for each pixel, a single value is calculated for the model by computing an average across the image, weighted by total final linear polarization of each pixel. 

\begin{figure}
  \centering
  \includegraphics[width=0.5\textwidth]{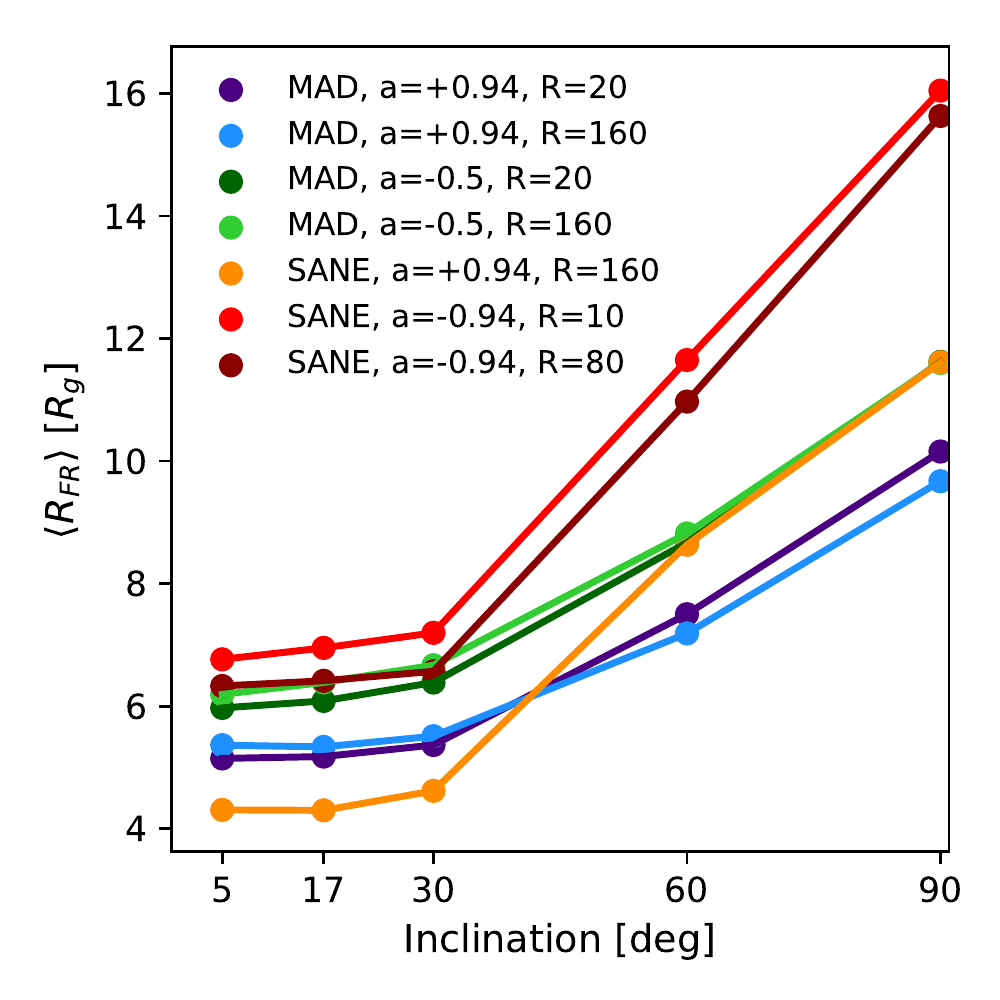}
  \caption{Characteristic distance of Faraday rotating material in these models during their final snapshot as a function of inclination.  Due to the evacuated jet region in these simulations, the Faraday rotating material is confined to low radius at low inclination, but extends to larger radius at higher inclination.  Recall that our calculations terminate at $R=20 \ G M_\bullet/c^2$, within which these simulations are in inflow equilibrium.  \label{fig:faradayRotationDistance}}
\end{figure}

In Figure \ref{fig:faradayRotationDistance}, we plot the characteristic distance of Faraday rotating material of these models during their final snapshot as a function of inclination.  For inclinations $<30^\circ$, most of the Faraday rotation occurs at $<10 \; GM_\bullet/c^2$, while $\langle R_{FR} \rangle$ increases at higher inclinations.  The innermost stable circular orbit exists at smaller radius for prograde models than for retrograde models, which leads to a noticeable difference in $\langle R_{FR} \rangle$ between these two classes of models at low inclination.

Recall that we restrict the domain of our calculations to within $20 \ G M_\bullet / c^2$, the radius within which the simulations are in inflow equilibrium.  As we further explore in Appendix \ref{sec:50M}, if this radius is increased to $50 M_\bullet$, we find consistent results for $i \leq 17^\circ$, but substantially larger $\langle R_{FR} \rangle$ for $i \geq 60^\circ$.  Therefore, we believe that our $|\mathrm{RM}|$ values throughout the paper should be considered lower limits for $i > 17^\circ$, as material from beyond the converged region may contribute to Faraday rotation at larger inclinations.

\begin{figure*}
  \centering
  \includegraphics[width=\textwidth]{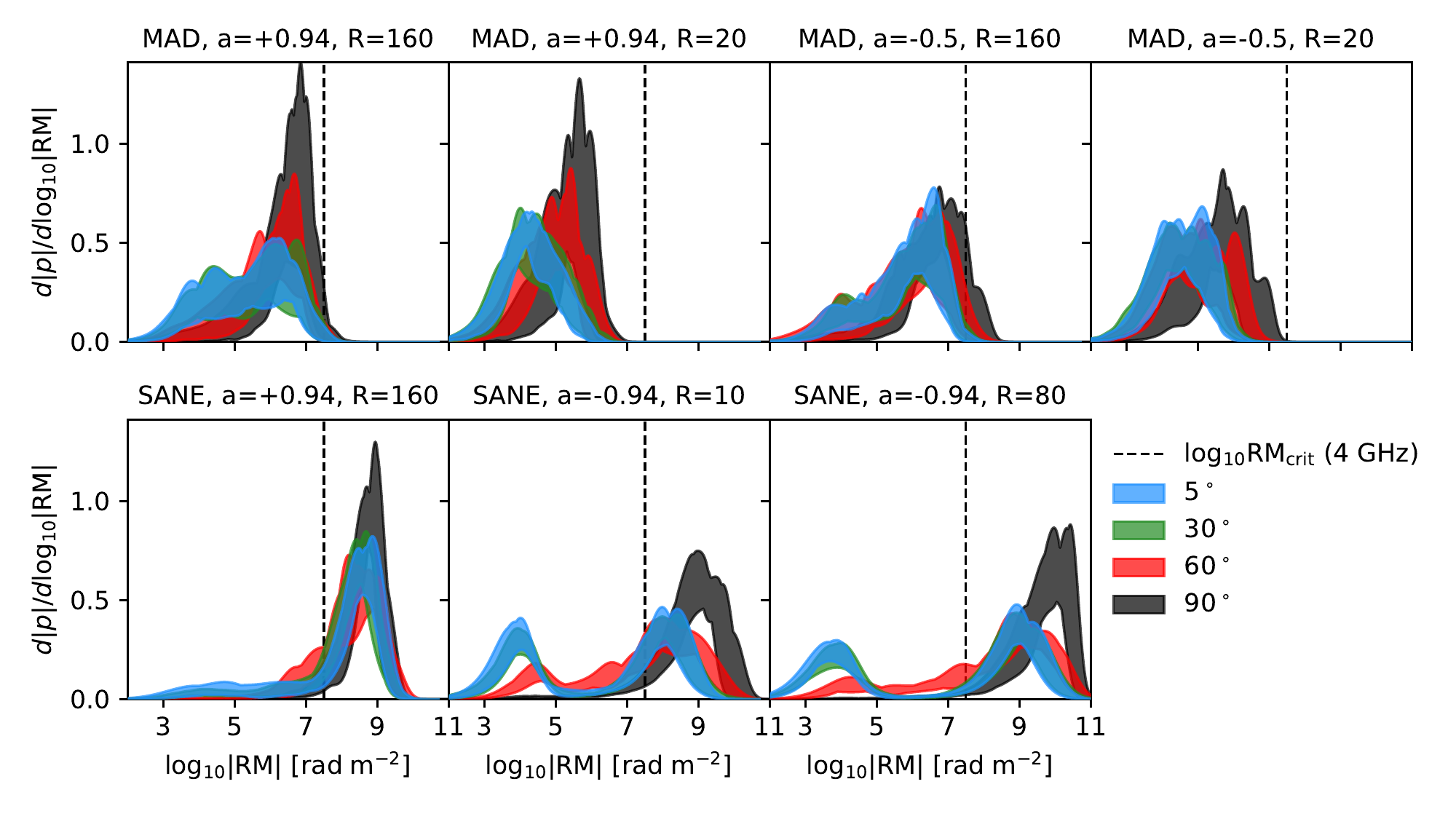}
  \caption{RM distribution functions as in Figure \ref{fig:rm_distribution_functions}, now shown as a function of inclination.  Both halves of the emitting region are combined in this figure.  At higher inclinations, we find that these distributions are skewed towards higher values, as the population of photons experiencing comparatively little RM diminishes.  \label{fig:rm_distributions_all}}
\end{figure*}

Since the RM is highly non-uniform across these images, spatially resolved RM distribution functions provide greater insight into the inclination dependence.  In Figure \ref{fig:rm_distributions_all}, we plot the RM distribution functions as in Figure \ref{fig:rm_distribution_functions}, for inclinations $i \in \{5^\circ,30^\circ,60^\circ,90^\circ\}$.  Unlike in Figure \ref{fig:rm_distribution_functions}, we do not split the distributions into their front and back halves.  At low inclination, these models exhibit significant emission with low $|\mathrm{RM}|$.  As the inclination approaches $90^\circ$, this fraction of emission with low $|\mathrm{RM}|$ diminishes, and the distribution is skewed towards higher values.  Note that the distributions at $5^\circ$ are indistinguishable from those at $30^\circ$.

\subsection{Non-Linear $\chi(\lambda^2)$ as a Model Discriminant}
\label{sec:nonlinearity}

\begin{figure*}
  \centering
  \includegraphics[width=\textwidth]{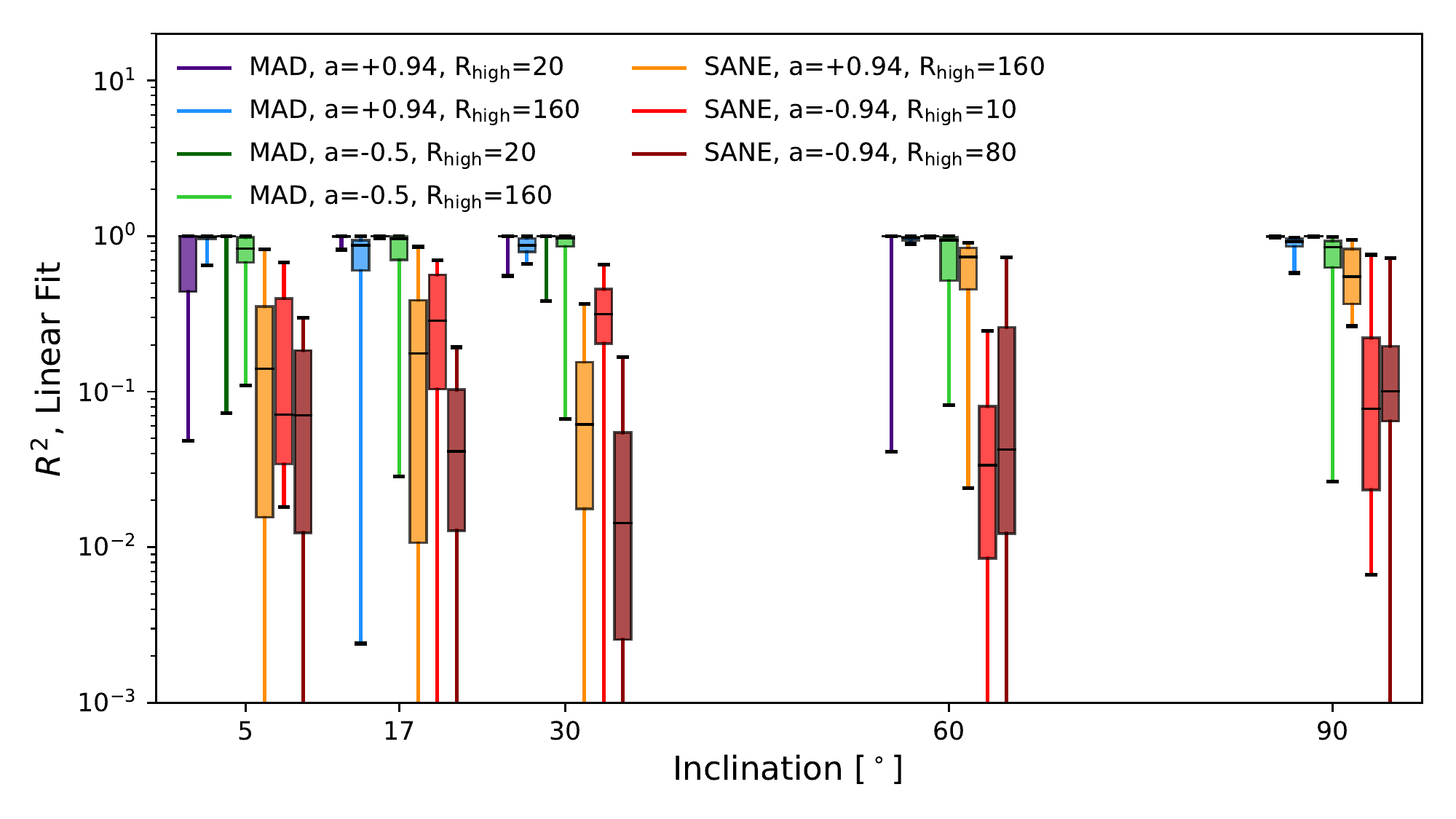}
  \caption{Non-linearity of $\chi(\lambda^2)$ among the models considered in this study.  $R^2$ of a linear fit to $\chi(\lambda^2)$ describes the fraction of variation in the data that can be explained by a simple linear model.  SANE models exhibit more non-linearity than MAD models in this study, since their images contain a significant amount of polarized intensity with $|\mathrm{RM}| > |\mathrm{RM}|_\mathrm{crit}$.  \label{fig:rm_nonLinearity}}
\end{figure*}

Since emission and Faraday rotation occur co-spatially and non-uniformly throughout these models, $\chi(\lambda^2)$ need not be linear.  We find that the degree of non-linearity varies significantly among models, due to the complex RM structure described in \S\ref{sec:spatial_variation}.  As a metric of non-linearity, we fit lines to $\chi(\lambda^2)$ from the 16 small bands spanning 226 to 230 GHz and obtain the coefficient of determination, $R^2$, defined via

\begin{equation}
    R^2 = 1 - \frac{SS_{res}}{SS_{tot}}.
\end{equation}

Here, $SS_{res}$ is the regression sum of squares, and $SS_{tot}$ is the total sum of squares.  $R^2$ describes the fraction of variation within the data that can be ascribed to the simple linear dependence.

Our results are shown in Figure \ref{fig:rm_nonLinearity}, following the same formatting as Figure \ref{fig:rm_inclination}.  Based on these results, SANE models should exhibit non-linear behaviour most of the time.  In contrast, MAD models are almost always well described by a linear $\chi(\lambda^2)$ law, especially those with $\mathrm{R}_\mathrm{high}=20$.  This can be understood by returning to Figure \ref{fig:rm_distribution_functions}.  Pixels with $|\mathrm{RM}| > |\mathrm{RM}|_\mathrm{crit}$ have individual EVPAs that rotate substantially across the 4 GHz bandwidth.  Images consisting of a substantial fraction of such pixels will therefore exhibit structure in $\chi(\lambda^2)$ within the bandwidth.  Among the models considered in this study, this behaviour appears much more likely among SANEs.

\subsection{Case Study:  RM Time Variability}
\label{sec:case_study}

We study one model with higher time resolution in order to quantify the variability of its RM.  For the MAD, $a=0.94$, $R_\mathrm{high}=20$ model,  we create images for every available snapshot within $t/(GM_\bullet/c^3) \in [7500,10000]$, which are each separated by $5 \; GM_\bullet/c^3$.  This model has the smallest accretion rate and $|\mathrm{RM}|$ of the models explored in this work.  Its $|\mathrm{RM}|$ is sufficiently smaller than $\mathrm{RM}_\mathrm{crit}$ that $\chi(\lambda^2)$ remains linear within the bandwidth (see Figure \ref{fig:nonlambda2_16bands}), and thus we create only 2 polarized images (two frequencies) at each snapshot instead of the usual 6 (three frequencies, each separately for two sides of the disc).

\begin{figure*}
  \centering
  \includegraphics[width=\textwidth]{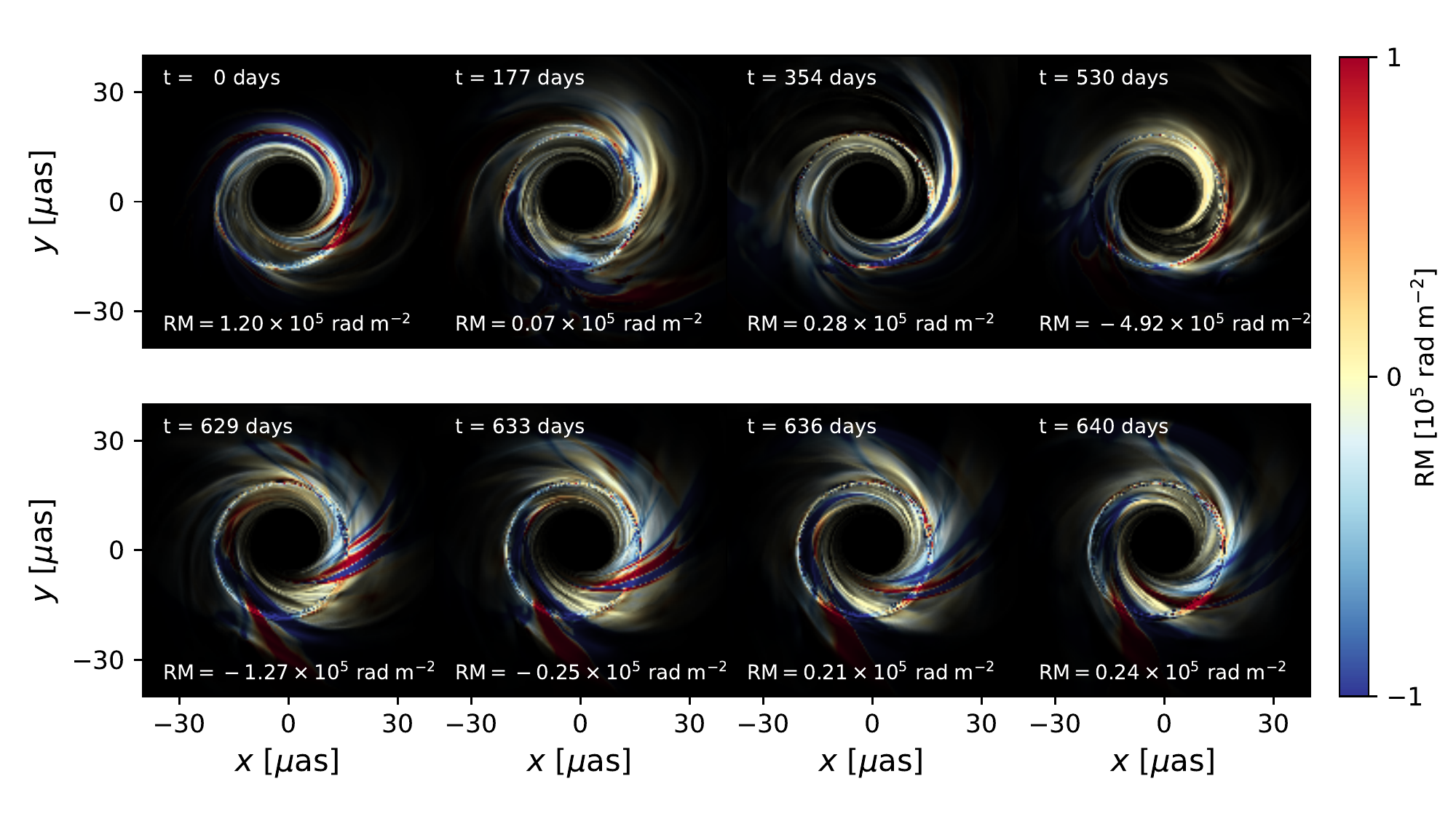}
  \caption{Visualisation of the time variability of the RM structure in the MAD, $a=0.94$, $R_\mathrm{high}=20$ model.  The brightness of each pixel scales with its linear polarized intensity, while the colour represents its rotation measure.  The RM that would be inferred from a spatially unresolved measurement is written at the bottom of each panel.  The spatially unresolved RM can change on time-scales of days as the Faraday rotating gas moves on event horizon scales.  The sign flip does not require a dramatic change in the source structure. \label{fig:time_series_visualisation}}
\end{figure*}

The time variability of this model at an inclination of $17^\circ$ is visualised in Figure \ref{fig:time_series_visualisation}.  In the top row, panels are separated by about half a year, while in the bottom row panels are separated by about 5 days.  As in previous figures, pixel brightness encodes the linear polarized intensity, while the colour encodes the RM.  Both positive and negative RM regions can be found in a typical snapshot.  The spatially unresolved RM is written at the bottom of each panel.  The bottom row illustrates how the dynamics of an image with both positive and negative RM regions can result in variability and RM sign flips on a timescale of a few days.  The RM sign flip does not require a drastic change in global source structure; rather, the balance between positive and negative RM regions is shifted.

\begin{figure*}
  \centering
  \includegraphics[width=\textwidth]{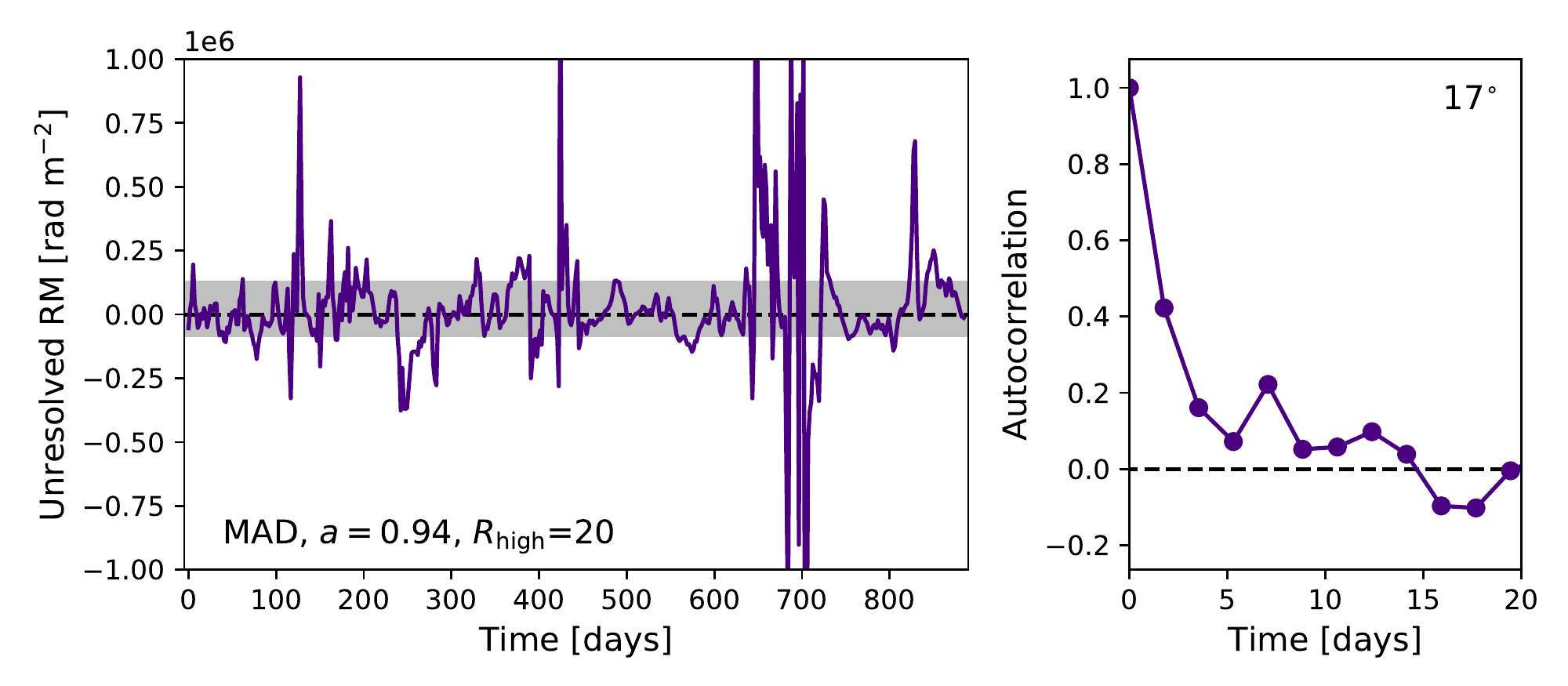}
  \caption{{\it Left:}  RM as a function of time in the MAD, $a=0.94$, $\mathrm{R}_\mathrm{high}=20$ model that we study with greater time resolution.  The grey band encloses the $1\sigma$ percentile region over this time, which includes both positive and negative values.  {\it Right:}  Auto-correlation function of this time series.  The auto-correlation drops below 50 per cent in less than the separation between snapshots. \label{fig:time_series}}
\end{figure*}

In Figure \ref{fig:time_series}, we plot the RM as a function of time for this model, as well as its auto-correlation function.  The geometrized time unit is converted to days via $t_\bullet = GM_\bullet/c^3$, which for M87* is 8.5 hours.  The grey band encloses the 16th to 84th ($1\sigma$) percentiles.  At $17^\circ$, these include both positive and negative values, such that $\mathrm{RM} = {-0.00}^{+1.27}_{-0.83} \times 10^5 \ \mathrm{rad} \; \mathrm{m}^{-2}$.  Examining the images, we do not notice any obvious special behaviour in the accretion flow during periods of large $|\mathrm{RM}|$.  Since the RM is determined by the motion of material on event horizon scales, the auto-correlation function of this time series drops rapidly, falling below 0.5 in less than the separation between snapshots.

\begin{figure*}
  \centering
  \includegraphics[width=\textwidth]{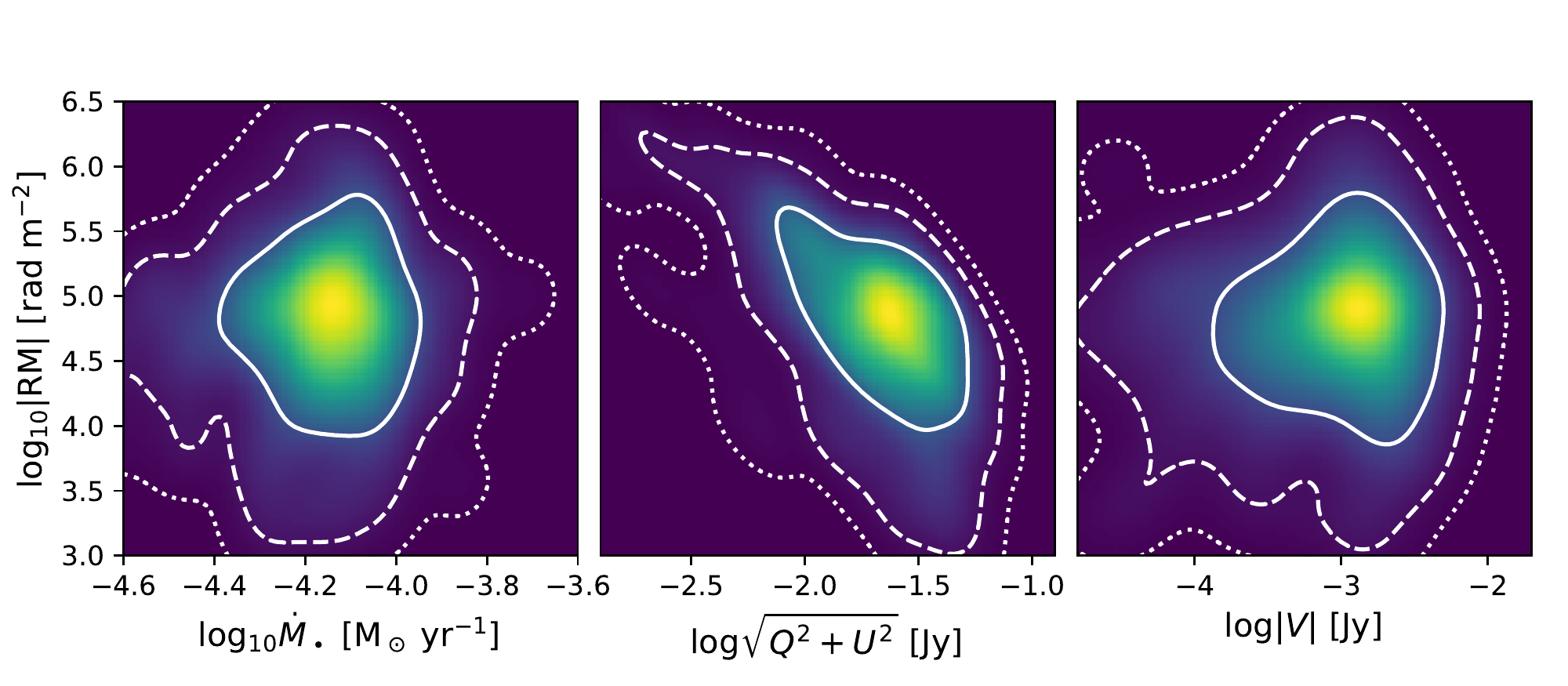}
  \caption{Joint probability distributions of $\log_{10}|RM|$ with accretion rate, linear polarized intensity, and circular polarized intensity in the MAD, $a=0.94$, $\mathrm{R}_\mathrm{high}=20$ model with an inclination of 17$^\circ$. One, two, and three sigma contours are overlaid in white, using solid, dashed, and dotted lines respectively.  We find no correlation between $|\mathrm{RM}|$ and $\dot{M}_\bullet$, implying that a change in RM does not imply a change in the accretion rate.  As observed by \citet{Bower+2018} for Sgr A*, we recover an anti-correlation between $|\mathrm{RM}|$ and linear polarization, but not circular polarization. \label{fig:probability_distributions}}
\end{figure*}

In Figure \ref{fig:probability_distributions}, we plot the joint probability distributions of $\log_{10}|RM|$ with accretion rate, linear polarized intensity, and circular polarized intensity for the model at $17^\circ$.  One, two, and three sigma contours are overlaid in white.  We do not find any correlation between $|\mathrm{RM}|$ and $\dot{M}_\bullet$, indicating that within a single model, a change in RM does not imply a change in the accretion rate, as might be suggested by analytic models.   Rather, as we have discussed, $|\mathrm{RM}|$ and its sign appears to result from a complicated and stochastic cancellation of positive and negative regions.  For Sgr A*, \citet{Bower+2018} found an anti-correlation between linear polarized intensity and RM, but no such correlation with circular polarized intensity.  For this particular model of M87*, we recover qualitatively similar results: a linear regression yields $\log_{10}|RM| = -1.2 \log_{10}LP + 2.7 \pm 0.08$, where $LP = \sqrt{Q^2+U^2}$ in Jy and $\mathrm{RM}$ is in units of rad m$^{-2}$, with a moderate r-value of -0.57.  No statistically significant correlation is found between $|\mathrm{RM}|$ and circular polarization.  An anti-correlation between $|\mathrm{RM}|$ and LP is not surprising, since greater Faraday rotation implies greater scrambling of the polarization vector field.

\section{Discussion}
\label{sec:discussion}

\subsection{Comparing Different Models}

Here, we summarize the qualitative commonalities and differences between the different models we have considered.  
\begin{itemize}
  \item {\bf Prograde MAD}:  These models require the lowest accretion rate to generate the appropriate total intensity, and consequently exhibit the lowest $|\mathrm{RM}|$.  Compared to the other models, there is not too much difference in $|\mathrm{RM}|$ between the two halves of the emitting region, since both components occur close to the mid-plane.  These models usually exhibit linear $\chi(\lambda^2)$ within 4 GHz.
  \item {\bf Retrograde MAD}: Retrograde MAD models require larger accretion rates than their prograde counterparts, but exhibit similar values of $|\mathrm{RM}|$.  Some areas of the $\mathrm{R}_\mathrm{high}=160$ models have large enough $|\mathrm{RM}|$ to weakly bandwidth depolarize or scramble portions of the image.
  \item{\bf Prograde SANE}:  Due to lensing, the counter-jet spans a larger angular scale and contributes more to the total intensity than the forward-jet.  Bandwidth depolarization effects are severe, and emission from the counter-jet is entirely depolarized, assuming a 4 GHz bandwidth.  Consequently, the total intensity image (dominated by the counter-jet) appears morphologically different compared to the linearly polarized image (dominated by the forward-jet).  $\chi(\lambda^2)$ exhibits strongly non-linear evolution.
  \item{\bf Retrograde SANE}:  The counter-jet experiences 6 orders of magnitude more Faraday rotation than the forward-jet.  As with their prograde counterparts, $\chi(\lambda^2)$ is highly non-linear.  However, the difference in morphology between the total intensity image and the linearly polarized image is less significant than the prograde case, since the two emission components subtend more similar angular scales.
\end{itemize}
By construction, increasing $R_\mathrm{high}$ decreases the temperature of electrons in the mid-plane, which therefore increases the Faraday depth for emission that passes through it.  In all models, larger $R_\mathrm{high}$ results in greater non-linearity of $\chi(\lambda^2)$.

\subsection{RM and Bandwidth Depolarization}

If anywhere in an image, $\mathrm{RM} > \mathrm{RM}_\mathrm{crit}$, then that region's linear polarization should be suppressed as described by equation \ref{eqn:bw_depolarization}.  We find that this is more likely to occur in SANE models, but may also affect some parts of MAD models.  Bandwidth depolarized regions manifest as areas with lower than average linear polarization fraction.

In some of the worst cases, like that presented in Figure \ref{fig:taylor_expansion_example}, we find that basic image properties are affected by bandwidth depolarization, such as the linear polarization fraction as well as the morphology of the linear polarization vector field, even after images are blurred.  This may be important for studies which use images computed at a single frequency to compare to observations taken over a finite bandwidth.  For future studies, the methodology introduced in \S\ref{sec:taylor_expansion} can be generalized by applying the Taylor expansion to each point along the ray-traced geodesic instead of just two separate emission regions.  This would allow the appropriate bandwidth integrations to occur within the ray-tracing code itself.

\subsection{Caveats and Limitations}

At present, the electron distribution function is poorly constrained.  For creating these images, a thermal distribution function is assumed, along with the $R_\mathrm{high}$ prescription developed by \citet{Moscibrodzka+2016}.  \citet{Mao+2017} studied the effects of adding a power-law component to the distribution function and found that even if a few percent of the total energy is put into a non-thermal power law component, a diffuse halo of emission can be produced.  The effects of non-thermal electron distribution functions on polarized images remain to be studied.

Recall that we truncate our radiative transfer calculations at a radius of $20 \ GM_\bullet/c^2$, only within which these GRMHD simulations exhibit inflow equilibrium.  However, Faraday rotating material can plausibly exist at larger radius, especially at higher inclinations.  Using very long duration simulations, \citet{Dexter+2020} find that Faraday rotation can peak at radii $R \sim 30-90 \; GM_\bullet/c^2$, depending on the electron prescriptions.  A more distant Faraday screen would be expected to uniformly rotate all EVPAs by a fixed amount, which may leave signatures in the EVPA vector field \citep{Palumbo+2020}.  Additionally, such a screen should maintain a consistent RM for longer timescales than these models.  Such distant screens may in fact be required to explain the consistent RMs of Sgr A* \citep{Bower+2018} and 3C 84 \citep{Plambeck+2014} over timescales of years.

\section{Summary and Conclusions}
\label{sec:conclusion}

We have investigated rotation measure (RM) within a subset of the EHT simulation library that is consistent with the observational constraints on M87* considered in \citetalias{EHT5}.  We find more information in the RM structure of GRMHD simulations than can be described by a single scalar, $\mathrm{RM}$.  We summarise our results below:

\begin{itemize}
    \item In a single snapshot, we find extreme variations in the RM between different regions.  The RM may vary by orders of magnitude and even flip sign across the image.  The RM inferred from a spatially unresolved measurement is therefore the result of the complicated interplay of these different regions.
    \item Emission originating from in front of the disk mid-plane may be orders of magnitude less Faraday rotated than emission from the back.  In the high accretion rate SANE models, the RM is large enough to completely depolarize emission from the counter-jet, which may in fact dominate the total intensity.  The sub-relativistic mid-plane is the dominant source of Faraday rotation in these models.
    \item Many models exhibit clear departures from a $\lambda^2$ law even across a narrow fractional bandwidth of 4 GHz.  Non-linearity is a more common feature among the SANE models, and increases with $R_\mathrm{high}$.
    \item The RM structure changes as material moves on event horizon scales.  These models all exhibit strong time variability, causing the spatially unresolved RM to vary and even flip sign on a time-scale of days. 
    \item RM is a poor predictor of the accretion rate.  These models predict several orders of magnitude spread in RM for a given accretion rate, and within a single model, these quantities are not correlated as a function of time.  In addition, analytic models used to infer the accretion rate based on the RM \citep{Marrone+2006} systematically underestimate the accretion rate onto M87*, since the source should be viewed through an evacuated funnel region.
\end{itemize}

In future work, a more thorough investigation of the EHT simulation library is merited, including models for Sgr A*.  Alternative models for the electron distribution function should also be considered.  Repeated observations of both Sgr A* and M87* will be useful to probe the time variability predicted by these models.

\section{acknowledgements}

We thank Jason Dexter for valuable feedback which greatly improved the content of this paper.  We also thank Laurent Loinard and the anonymous referee for their careful and thorough reviews.  This material is based upon work supported by the National Science Foundation under Grant No. OISE 1743747 and NSF AST 1716327.  Computations were performed using the resources of the Black Hole Initiative (BHI).  Computations at the BHI were made possible through the support of grants from the Gordon and Betty Moore Foundation and the John Templeton Foundation.  This work used the Extreme Science and Engineering Discovery Environment (XSEDE) resource stampede2 at TACC through allocation TG-AST170024.
The opinions expressed in this publication are those of the author(s) and do not necessarily reflect the views of the Moore or Templeton Foundations.

\section{Data availability}

The data underlying this article will be shared on reasonable request to the corresponding author.

\bibliography{ms}

\appendix

\section{Spherical Stokes Parameters}
\label{sec:spherical_stokes}

In \S\ref{sec:taylor_expansion}, we convert the standard Stokes parameters $\{Q,U,V\}$ to the spherical Stokes parameters $\{N,\phi,\psi\}$ as in \citet{Shcherbakov+2012} for the purposes of a more stable Taylor expansion.  This transformation is defined by:
\begin{align}
    Q &= N\cos\phi\sin\psi \\
    U &= N\sin\phi\sin\psi \\
    V &= N\cos\psi
\end{align}
while its inversion can be derived from trigonometric identities as
\begin{align} 
    N &= \sqrt{Q^2+U^2+V^2} \\
    \phi &= \arctan(U/Q) \\
    \psi &= \arctan(\sqrt{Q^2+U^2} / V).
\end{align}
From these equations, we can see that $N$ is the total amount of both linear and circular polarization, $\phi$ describes the phase of the linear polarization, and $\psi$ describes the linear to circular polarization ratio.  Note the lack of a factor of 1/2 in $\phi$, such that $\phi = 2\chi$, where $\chi$ is the EVPA.

In a pixel with a large $|\mathrm{RM}|$, Faraday rotation causes $Q$ and $U$ to oscillate rapidly with frequency.  In spherical Stokes parameters, $N$ remains stable, while $\phi$ changes linearly with wavelength squared, with $d\phi/d\lambda^2 = 2\mathrm{RM}$.  This makes the Spherical stokes parameters more stable to Taylor expansion.

When we calculate derivatives of the $\{N,\phi,\psi\}$, we compute them in terms of $\{Q,U,V\}$ and their derivatives.  This allows us to avoid mistakes due to phase wrapping, as discussed in \S\ref{sec:taylor_expansion}.  By simply differentiating according to the chain rule, the derivatives are given by
\begin{align}
    N' &= \frac{QQ' + UU' + VV'}{\sqrt{Q^2+U^2+V^2}} \\
    \phi' &= \frac{U'Q -Q'U}{Q^2+U^2}\\
    \psi' &= \frac{V(QQ'+UU') - V'(Q^2+U^2)}{(Q^2 + U^2 + V^2)\sqrt{Q^2+U^2}}
\end{align}
where $'$ denotes differentiation with respect to frequency (or another physically similar quantity such as wavelength or wavelength-squared).

\section{Analytic Bandwidth Integrals}
\label{sec:analyticIntegrals}

Here, we provide analytic solutions to the integrals described in \S\ref{sec:taylor_expansion}.  Spherical Stokes parameters $\{I_0, N_0, \phi_0, \psi_0\}$ and their derivatives $\{dI/d\nu, dN/d\nu, d\phi/d\lambda^2, d\psi/d\lambda^3\}$ are estimated at frequency $\nu_0$.  Let $\nu_c$ be the central frequency of a band extending between $\nu_1 = \nu_c-\Delta\nu/2$ and $\nu_2 = \nu_c + \Delta\nu/2$.  We then define a dimensionless frequency $x = (\nu-\nu_0)/\Delta\nu$ such that $x_1 = x(\nu_1)$ and $x_2 = x(\nu_2)$.  We can then write

\begin{align}
    I_\mathrm{BW} &= \int_{x_1}^{x_2} (I_0 + I_1x) dx \\
    Q_\mathrm{BW} &= \int_{x_1}^{x_2} (N_0 + N_1x) \cos(\phi_0 + \phi_1 x) \sin (\psi_0 + \psi_1 x) dx \\
    U_\mathrm{BW} &= \int_{x_1}^{x_2} (N_0 + N_1x) \sin(\phi_0 + \phi_1 x) \sin (\psi_0 + \psi_1 x) dx \\
    V_\mathrm{BW} &= \int_{x_1}^{x_2} (N_0 + N_1x) \cos (\psi_0 + \psi_1 x) dx
\end{align}

\noindent where $x_1 = (\nu_c-\nu_0-\Delta\nu/2)/\Delta\nu$, $x_2 = (\nu_c-\nu_0+\Delta\nu/2)/\Delta\nu$, and we define the following quantities from the derivatives of the spherical Stokes parameters:

\begin{align}
    I_1 &= \frac{dI}{d\nu}\Delta\nu \\
    N_1 &= \frac{dN}{d\nu}\Delta\nu \\
    \phi_1 &= \frac{-2c^2\Delta\nu}{\nu_0^3}\frac{d\phi}{d\lambda^2} \\
    \psi_1 &= \frac{-3c^3\Delta\nu}{\nu_0^4}\frac{d\psi}{d\lambda^3}
\end{align}

These integrals have analytic solutions given by

\begin{equation}
    I_\mathrm{BW} = I_0(x_2 - x_1) + \frac{I_1}{2}\Big(x_2^2 - x_1^2\Big)
\end{equation}

\begin{equation}
    \begin{aligned}
    &Q_\mathrm{BW} = \frac{1}{2} \Big[N_0 \Big(\frac{\cos (\phi_0+\phi_1
   x_2-\psi_0-\psi_1 x_2)}{\phi_1-\psi_1}-\\&\frac{\cos (\phi_0+\phi_1
   x_1-\psi_0-\psi_1 x_1)}{\phi_1-\psi_1}+\\
   &\frac{\cos
   (\phi_0+x_1 (\phi_1+\psi_1)+\psi_0}{\phi_1+\psi_1})\\&-\frac{\cos
   (\phi_0+x_2
   (\phi_1+\psi_1)+\psi_0)}{\phi_1+\psi_1}\Big)+\\
   &N_1\Big(\frac{\sin (\phi_0+\phi_1 x_1-\psi_0-\psi_1
   x_1)}{(\phi_1-\psi_1)^2}-\\&\frac{x_1 (\phi_1-\psi_1) \cos (\phi_0+\phi_1
   x_1-\psi_0-\psi_1
   x_1)}{(\phi_1-\psi_1)^2}+\\&\frac{x_1 (\phi_1+\psi_1)
   \cos (\phi_0+x_1 (\phi_1+\psi_1)+\psi_0)}{(\phi_1+\psi_1)^2}\\&-\frac{\sin
   (\phi_0+x_1
   (\phi_1+\psi_1)+\psi_0)}{(\phi_1+\psi_1)^2}+\\&\frac{x_2
   (\phi_1-\psi_1) \cos (\phi_0+\phi_1
   x_2-\psi_0-\psi_1 x_2)}{(\phi_1-\psi_1)^2}-\\&\frac{\sin (\phi_0+\phi_1
   x_2-\psi_0-\psi_1 x_2)}{(\phi_1-\psi_1)^2}+\\&\frac{\sin
   (\phi_0+x_2 (\phi_1+\psi_1)+\psi_0)}{(\phi_1+\psi_1)^2}-\\&\frac{x_2
   (\phi_1+\psi_1) \cos (\phi_0+x_2
   (\phi_1+\psi_1)+\psi_0)}{(\phi_1+\psi_1)^2}\Big)\Big]
   \end{aligned}
\end{equation}

\begin{equation}
    \begin{split}
    &U_\mathrm{BW} = \frac{1}{2} N_0 \Big[-\frac{\sin (\phi_0+\phi_1
   x_1-\psi_0-\psi_1 x_1)}{\phi_1-\psi_1}+\\&\frac{\sin
   (\phi_0+x_1
   (\phi_1+\psi_1)+\psi_0)}{\phi_1+\psi_1}+\\&\frac{\sin
   (\phi_0+\phi_1 x_2-\psi_0-\psi_1
   x_2)}{\phi_1-\psi_1}-\\&\frac{\sin (\phi_0+x_2
   (\phi_1+\psi_1)+\psi_0)}{\phi_1+\psi_1}\Big]+\\&\frac{1}{2}
   N_1 \Big[-\frac{x_1 (\phi_1-\psi_1) \sin
   (\phi_0+\phi_1 x_1-\psi_0-\psi_1 x_1)}{(\phi_1-\psi_1)^2}+\\& \frac{\cos
   (\phi_0+\phi_1 x_1-\psi_0-\psi_1
   x_1)}{(\phi_1-\psi_1)^2}+\\& \frac{x_1 (\phi_1+\psi_1)
   \sin (\phi_0+x_1 (\phi_1+\psi_1)+\psi_0)}{(\phi_1+\psi_1)^2}+\\&\frac{\cos
   (\phi_0+x_1
   (\phi_1+\psi_1)+\psi_0)}{(\phi_1+\psi_1)^2}+\\&\frac{x_2
   (\phi_1-\psi_1) \sin (\phi_0+\phi_1
   x_2-\psi_0-\psi_1 x_2)}{(\phi_1-\psi_1)^2}+\\&\frac{\cos (\phi_0+\phi_1
   x_2-\psi_0-\psi_1
   x_2)}{(\phi_1-\psi_1)^2}-\\&\frac{x_2 (\phi_1+\psi_1)
   \sin (\phi_0+x_2 (\phi_1+\psi_1)+\psi_0)}{(\phi_1+\psi_1)^2}+\\& \frac{\cos
   (\phi_0+x_2
   (\phi_1+\psi_1)+\psi_0)}{(\phi_1+\psi_1)^2}\Big]
   \end{split}
\end{equation}

\begin{equation}
    \begin{aligned}
    &V_\mathrm{BW} = \frac{1}{\psi_1^2}\Big[-\psi_1 (N_0+N_1 x_1) \sin  (\psi_0+\psi_1 x_1)+ \\ 
    &\psi_1 (N_0+N_1 x_2) \sin (\psi_0+\psi_1 x_2)-\\
     &N_1 \cos (\psi_0+\psi_1 x_1)+N_1 \cos(\psi_0+\psi_1 x_2) \Big]
    \end{aligned}
\end{equation}

\section{Effects of Changing the Maximum Integration Radius}
\label{sec:50M}

In {\tt ipole}, the parameter {\tt rmax\_geo} (henceforth $R_\mathrm{out}$) sets the radius in Boyer-Lindquist (or Kerr-Schild) coordinates within which radiative transfer coefficients are calculated.  Although the MAD and SANE simulations have outer boundaries of $10^3 \ GM_\bullet/c^2$ and $50 \ GM_\bullet/c^2$ respectively, we find that these simulations exhibit inflow equilibrium only within $20 \ GM_\bullet/c^2$.  Hence, for the results throughout this paper, $R_\mathrm{out}$ is therefore set to $20 \ GM_\bullet/c^2$.  Here, we explore the effects of changing this outer radius to $50 \ GM_\bullet/c^2$.  Although this choice now includes material from unconverged regions, this allows us to gain some insight into how gas in more distant regions might affect our predictions.

\begin{figure*}
  \centering
  \includegraphics[width=\textwidth]{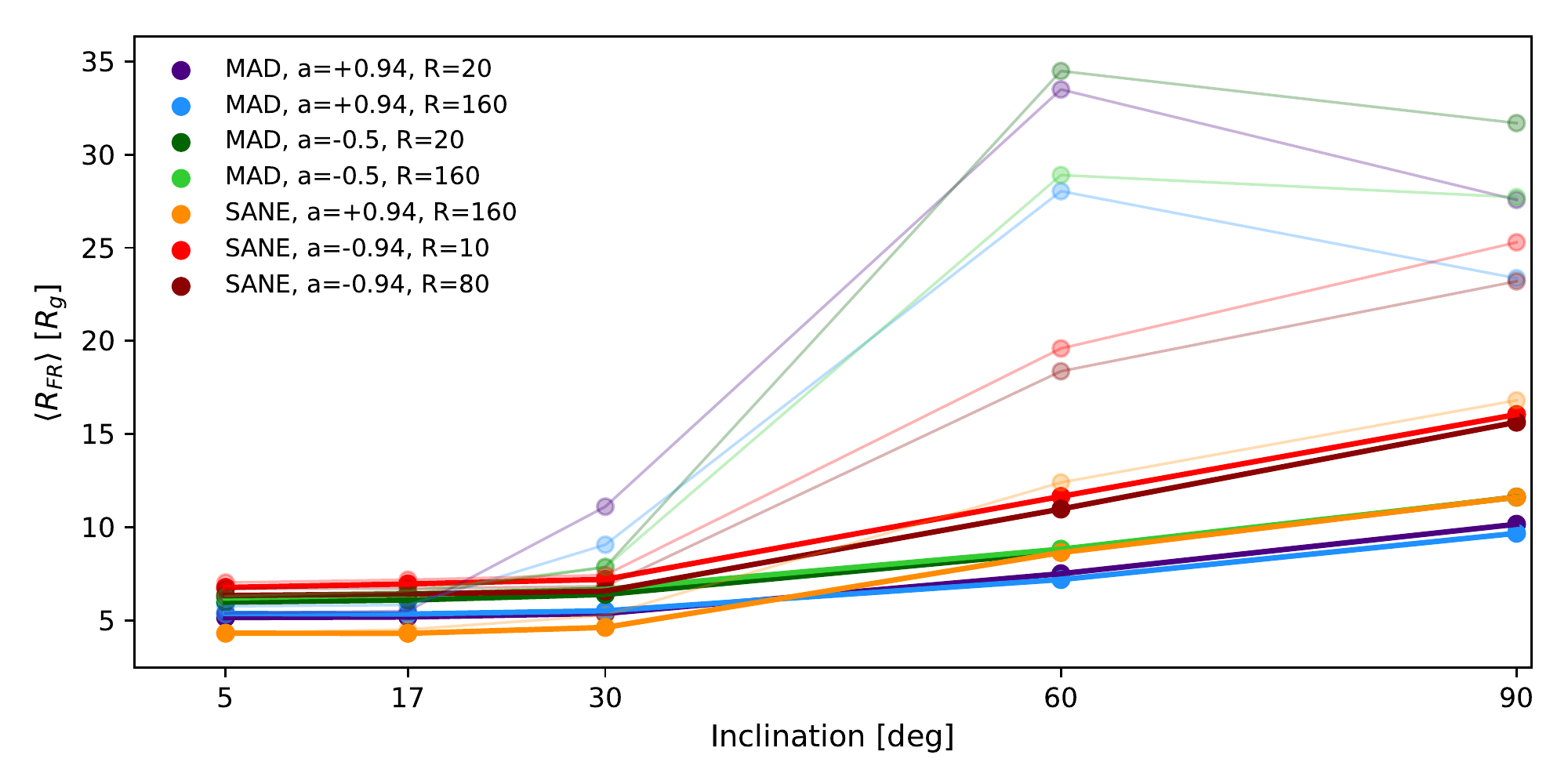}
  \caption{Characteristic distance of Faraday rotating material, as in Figure \ref{fig:faradayRotationDistance}, for $R_\mathrm{out} = 20 \ GM_\bullet/c^2$ models as solid lines, and $R_\mathrm{out} = 50 \ GM_\bullet/c^2$ models as faint lines.  There is little difference for inclinations $i \leq 17^\circ$, where the BH is viewed through an evacuated funnel region.  In contrast, material from more distant, unconverged regions can dominate the Faraday rotation at higher inclinations if calculations are allowed to proceed into this area.  \label{fig:faradayDistanceComparison}}
\end{figure*}

In Figure \ref{fig:faradayDistanceComparison}, we compare the characteristic distance of Faraday rotating material, $\langle R_{FR} \rangle$, as in Figure \ref{fig:faradayRotationDistance}.  The bold solid lines originate from the $R_\mathrm{out} = 20 \ GM_\bullet/c^2$ models (as shown in Figure \ref{fig:faradayRotationDistance}), while the faint thin lines originate from the $R_\mathrm{out} = 50 \ GM_\bullet/c^2$ models.  Interestingly, there is little difference in $\langle R_{FR} \rangle$ for inclinations $i\leq 17^\circ$, an evacuated funnel region in these simulations.  At larger inclinations, distant material from unconverged regions could potentially contribute the majority of the Faraday rotation.

\begin{figure*}
  \centering
  \includegraphics[width=\textwidth]{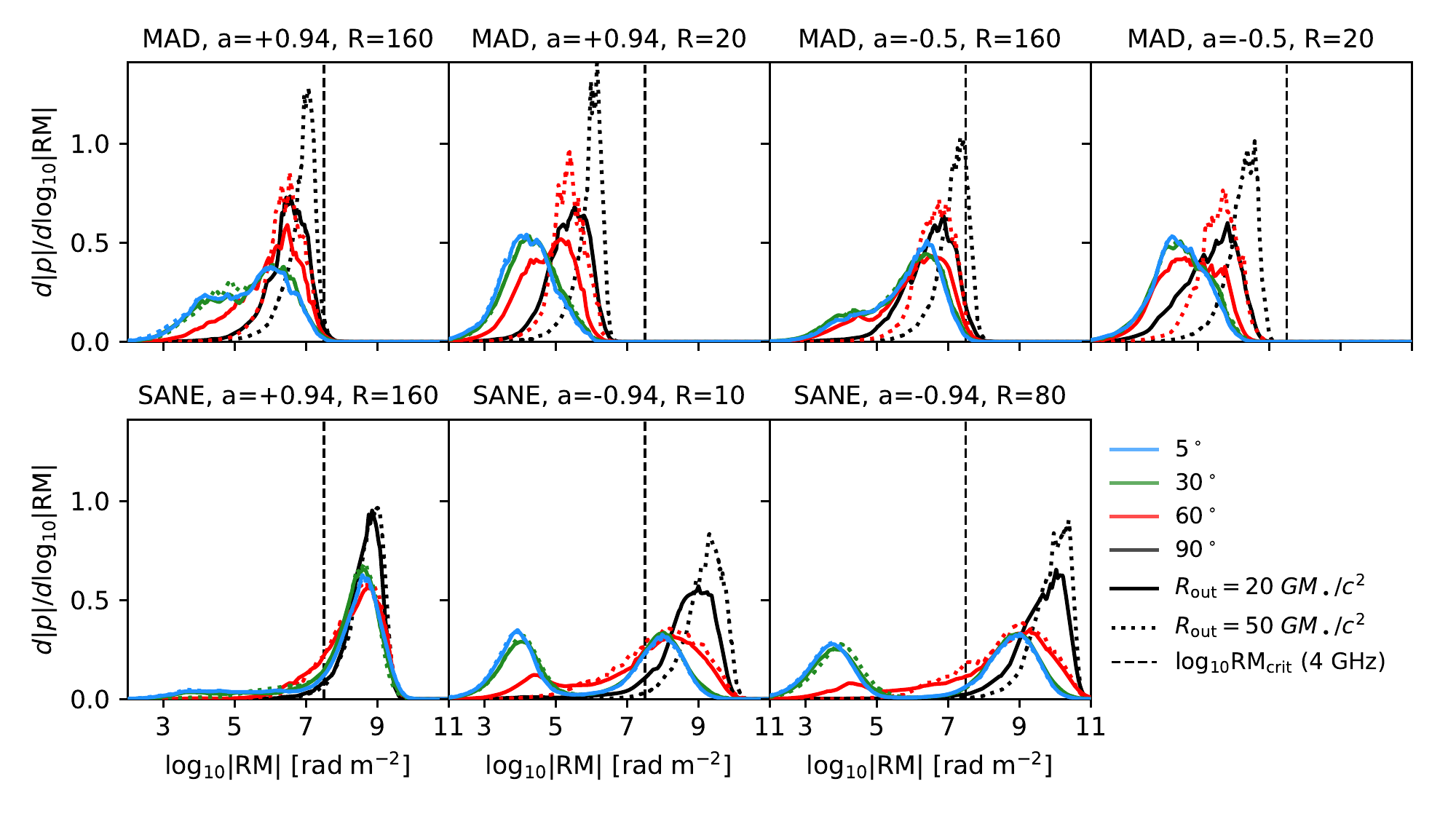}
  \caption{Rotation measure distribution functions, as in Figure \ref{fig:rm_distribution_functions}, where $R_\mathrm{out} = 20 \ GM_\bullet/c^2$ models are shown as solid lines and $R_\mathrm{out} = 50 \ GM_\bullet/c^2$ models are shown as dotted lines.  For clarity, only median values at a given $\log_{10}|\mathrm{RM}|$ are plotted.  There is negligible difference for all models when the inclination $\leq 30^\circ$.  The distributions are skewed towards higher values at $90^\circ$ for the retrograde SANEs, and for $\geq 60^\circ$ for MADs.  \label{fig:rm_distributions_comparison}}
\end{figure*}

In Figure \ref{fig:rm_distributions_comparison}, we plot the RM distribution functions as in Figure \ref{fig:rm_distributions_all}, where our $R_\mathrm{out} = 20 \ GM_\bullet/c^2$ results are shown as solid lines and alternative $R_\mathrm{out} = 50 \ GM_\bullet/c^2$ results are plotted as dotted lines.  For clarity, we only plot the median values at a given $\log_{10} |\mathrm{RM}|$ instead of the full range plotted in Figure \ref{fig:rm_distributions_all}.  We find that there is negligible difference in our results for inclinations $i \leq 30^\circ$, the range relevant for M87*.  At higher inclinations, the MAD distributions are skewed towards higher values for $i \geq 60^\circ$, while the retrograde SANE models only differ at $i=90^\circ$.  The prograde SANE model shows negligible difference between $R_\mathrm{out} = 20 \ GM_\bullet/c^2$ and $R_\mathrm{out} = 50 \ GM_\bullet/c^2$ even at $i=90^\circ$.  

\begin{figure*}
  \centering
  \includegraphics[width=\textwidth]{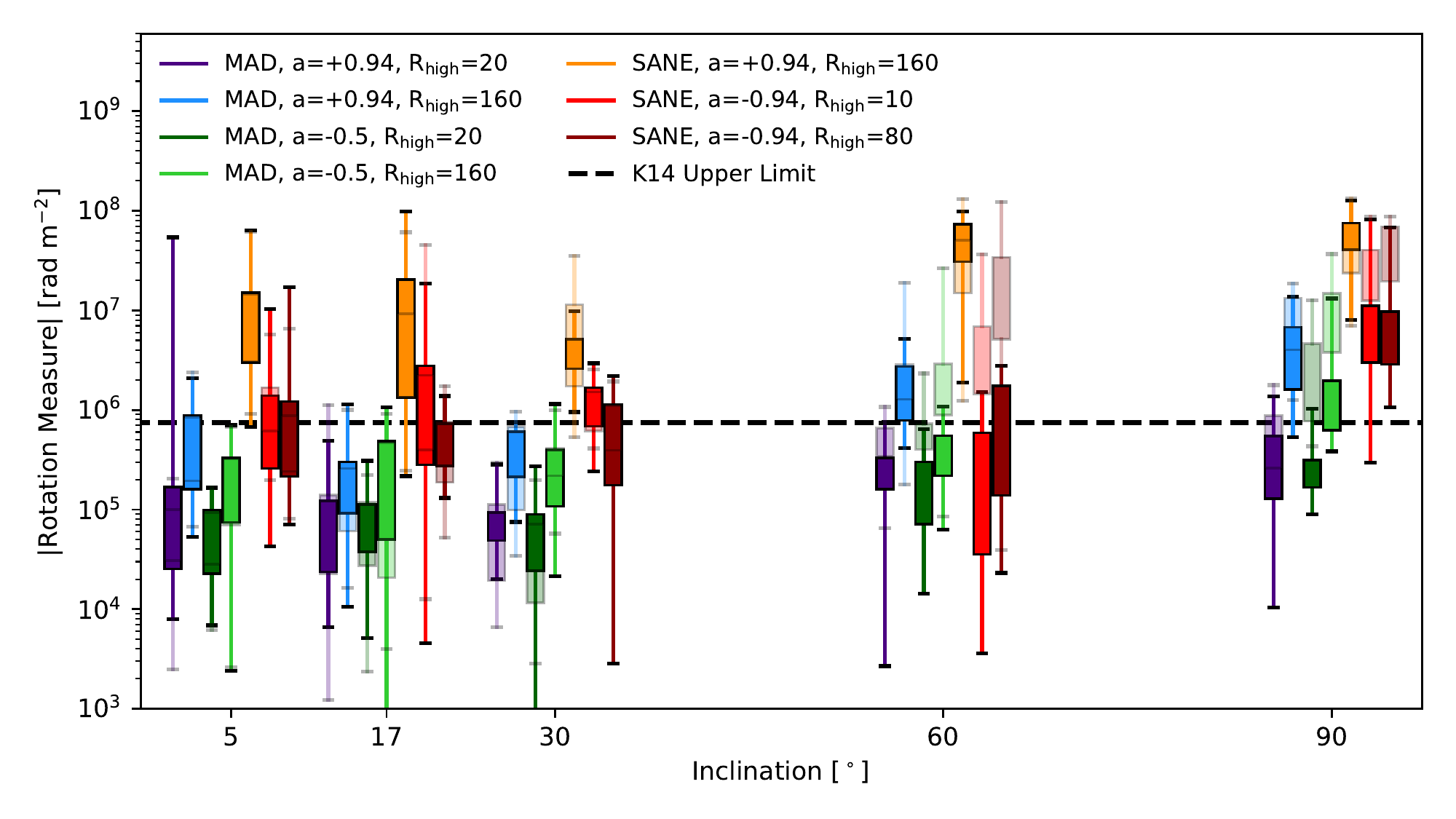}
  \caption{Rotation measure as a function of inclination, as in Figure \ref{fig:rm_inclination}, for $R_\mathrm{out} = 20 \ GM_\bullet/c^2$ models as solid boxes, and $R_\mathrm{out} = 50 \ GM_\bullet/c^2$ models as faint boxes.  While including material at $R_\mathrm{out}>20 \ GM_\bullet/c^2$ makes little difference for inclinations $i \leq 30^\circ$, it may increase the RM by factors of a few to orders of magnitude at larger inclinations, depending on the model.  \label{fig:rm_inclination_comparison}}
\end{figure*}

In Figure \ref{fig:rm_inclination_comparison}, we plot the effect this has on the spatially unresolved RM observed for these sources. Here, the solid boxes correspond to the $R_\mathrm{out} = 20 \ GM_\bullet/c^2$ models (as in Figure \ref{fig:rm_inclination}), while the faint boxes correspond to $R_\mathrm{out} = 50 \ GM_\bullet/c^2$ models.  Lines demarcating medians have been removed for clarity.  As expected, there is no noticeable difference for inclinations $i \leq 30^\circ$.  For inclinations of $\geq 60^\circ$, the RMs of $R_\mathrm{out} = 50 \ GM_\bullet/c^2$ models can be a factor of a few to orders of magnitude larger, depending on the model.

Finally, we notice that at inclinations of $60^\circ$, RM sign flips occur less frequently for the $R_\mathrm{out} = 50 \ GM_\bullet/c^2$ models than for the $R_\mathrm{out} = 20 \ GM_\bullet/c^2$ models.  This is to be expected, since material at larger radii evolves on longer timescales.

\end{document}